# Hydrogen diffusion in garnet: insights from atomistic simulations

Xin Zhong[1]; Felix Höfling[2,3]; Timm John[1]

[1]Freie Universität Berlin, Institut für Geologische Wissenschaften, Malteserstr. 74-100, 12249 Berlin, Germany

[2]Freie Universität Berlin, Fachbereich Mathematik und Informatik, Arnimallee 6, 14195 Berlin, Germany

[3]Zuse Institute Berlin, Takustr. 7, 14195 Berlin, Germany

**Abstract**

Garnet has been widely used to decipher the pressure-temperature-time history of rocks, but its physical properties such as elasticity and diffusion are strongly affected by trace amounts of hydrogen. Experimental measurements of H diffusion in garnet are limited to room pressure. We use atomistic simulations to study H diffusion in perfect and defective garnet lattices, focusing on protonation defects at the Si and Mg sites, which are shown to be energetically favored. The ab-initio simulation of H diffusion is computationally challenging due to a transient trapping of H, which is overcome with machine learning (ML) techniques by training a deep neural network that encodes the interatomic potential (IP). Our results show high mobility of hydrogen in defect free garnet lattices, whereas H diffusivity is significantly diminished in defective lattices. Tracer simulation focusing on H alone highlights the vital role of atomic vibrations of heavier atoms like Mg in the untrapping of H atoms. Two regimes of H diffusion are identified: a diffuser-dominated regime at high hydrogen content with low activation energies, due to saturation of vacancies by hydrogen, and a vacancy-dominated regime at low hydrogen content with high activation energies, due to trapping of H atoms at vacancy sites. These regimes account for experimental observations, such as a H-concentration dependent diffusivity and the discrepancy in activation energy between deprotonation and D-H exchange experiments. This study underpins the crucial role of vacancies in H diffusion and demonstrates the utility of machine-learned interatomic potentials in studying kinetic processes in the Earth's interior.

## Plain language summary

Garnet, though an anhydrous mineral in the Earth's crust and upper mantle, can contain trace amounts of hydrogen that may influence its physical properties. In this study, we investigated hydrogen diffusion in garnet using ab-initio molecular dynamics and classical molecular dynamics with a machine-learning interatomic potential. The machine-learning interatomic potential encodes the interatomic interaction via a deep neural network. Our results reveal that vacancies significantly impact hydrogen diffusivity in garnet. We identified two diffusion regimes: one dominated by the movement of interstitial hydrogen atoms and the other by cation vacancies at Mg and Si sites. Additionally, the atomic vibrations of heavier atoms like Mg, Si, and Al strongly influence the trapping and untrapping of hydrogen at high temperatures, thus affecting the overall diffusivity. These findings explain some experimental observations, such as hydrogen concentration-dependent diffusivity and the discrepancies in activation energy between deuterium-hydrogen exchange and deprotonation experiments. We also discussed the benefits and limitations of using machine-learning interatomic potentials in simulating kinetic processes within the Earth's interior.

# 1 Introduction

Garnet is a rock-forming anhydrous silicate mineral in the Earth's crust and upper mantle, whose formation is caused mainly by metamorphic reactions at elevated pressures and temperatures. Garnet may also form during low-temperature hydrothermal processes as well as magmatic crystallization (e.g. Chiama et al., 2023; Jamtveit and Hervig, 1994). It belongs to the nesosilicates and builds a complex solid solution covering a large chemical variety (Figure 1). While garnets formed in the crust are usually Fe-rich, mantle garnets are characterized by Mg-rich compositions, with both cations having similar ionic radii in the crystal lattice. Calcium is also commonly incorporated and results in an endmember with a comparably large molar volume. In particular, the formation of crustal garnet is associated with dehydration reactions, during which mineral-bound $H_2O$ is liberated into a dynamically forming porosity (Farber et al., 2014). Thus, garnet, even though being a nominally anhydrous mineral (NAM), may incorporate various amounts of water during formation or subsequent fluid influx (e.g. Lager et al., 1989; Rossman and Aines, 1991; Kaatz et al. 2022).

Water, often bound in the form of hydroxyl groups (OH) at lattice defects, can influence the physical properties of silicate minerals, such as their thermoelasticity (Jacobsen, 2018), electrical conductivity (Dai and Karato, 2009) or plastic strength (Xu et al., 2013; Zhong et al., 2024), as well as chemical processes, such as diffusion (e.g. Zhang et al., 2019). Owing to their small size, hydrogen atoms may easily migrate in the lattice of silicate minerals and combine with oxygen. If water is released, e.g., via the breakdown of hydrous minerals in subduction zones or transported upward or laterally via localized fluid pathways (John et al., 2012; Plümper et al., 2017), the physical properties of local rock may be substantially affected, and metamorphic reactions may be accelerated (Jamtveit et al., 2016; Kaatz et al., 2022; Putnis and John, 2010). Therefore, the migration of hydrogen in garnets may play a vital role in the dynamics of the lithosphere and asthenosphere.

The diffusion of hydrogen in garnet has been studied experimentally, e.g., via deprotonation experiments on H-bearing grossular and spessartine (Kurka et al., 2005; Phichaikamjornwut et al., 2012; Reynes et al., 2018) and deuterium–hydrogen (D–H) replacement experiments (Blanchard and Ingrin, 2004; Kurka et al., 2005). These experiments were performed at room pressure and at high temperatures of up to ca. 1400 K; however, hydrogen diffusion at high pressure was not quantified. Also, the formation energies of various point defects, such as Schottky and Frenkel defects, and their effect on the H diffusion coefficient are not fully understood. In this situation, *ab initio* calculations and molecular dynamics (MD) simulations can be helpful in determining the kinetics and fundamental mechanism of H diffusion.

Experiments on H incorporation and diffusion were performed for mantle minerals such as wadsleyite (Sun et al., 2021), and forsterite (Padrón-Navarta et al., 2014; Xue et al., 2017). *Ab initio* MD studies have been performed on mantle minerals, such as wadsleyite, ringwoodite (Caracas and Panero, 2017), and SrTiO perovskite (Shimojo and Hoshino, 2001), under high pressure–temperature (P–T) conditions. In recent studies, machine-learned interatomic potentials (MLIPs) have been used for the simulation of hydrogen diffusion in bridgmanite and post-perovskite (Peng and Deng, 2024) and for Li diffusion in melts (Luo et al., 2021). However, the application of MLIP to geomaterials is still in rapid progress, and its limitations and accuracy require exploration and cross-comparison with other complementary methods (Pinheiro et al., 2021).

## 2 Methods

Computational materials research, to a large extent, relies on the Born–Oppenheimer approximation, i.e., the separation of the dynamics of atomic electrons and nuclei. In MD simulations, the nuclei are treated as classical point particles and their many-body dynamics is generated by the classical (Hamiltonian) equations of motion. To this end, the forces on the nuclei stem from either empirical force fields or first-principles electronic structure calculations; the latter approach is referred to as *ab initio* MD (AIMD) simulations and is pursued in this work. We use density functional theory (DFT) to approximately solve the quantum mechanical many-electron problem for prescribed positions of the nuclei (Kohn and Sham, 1965). DFT is known to yield accurate predictions of many chemical and physical material properties (Sholl and Steckel, 2009). Next to the AIMD simulations, we use it to determine the formation energy of some point defects in garnet lattices, which informs us about the preferential locations of H atoms in the lattice. We note that treating the nuclei as classical particles, such as in AIMD, cannot yield the quantization of phonons and tunneling processes of the nuclei, to name a few potential limitations.

For the diffusion study, we combine three different approaches: 1) Car–Parrinello molecular dynamics (CPMD) simulations (Kühne, 2014; Marx and Hutter, 2010), which are computationally demanding since the DFT functional needs to be iterated in every integration timestep; 2) so-called tracer-MD simulations, which generate the motion of individual H atoms on a potential energy surface that is fixed in time and pre-computed via DFT; and 3) MLIP-MD simulations, which employ realistic interatomic potentials obtained from machine learning techniques with training data from DFT calculations.

Pyrope garnet has the composition $Mg_3Al_2Si_3O_{12}$ and occurs in a bcc structure with 80 atoms in the primitive unit cell, which can be extended to a cubic unit cell containing 160 atoms (Figure 1). The primitive unit cell was used for the CPMD simulations, whereas the cubic unit cell was used for the tracer-MD and MLIP-MD simulations.

## 2.1 DFT calculations and their validation

The DFT calculations are performed within the generalized gradient approximation (GGA) (Perdew et al., 1996) as implemented in the Quantum ESPRESSO (QE) software (Giannozzi et al., 2017, 2009), with the plane-wave cutoff energy set to 90 Ry. The optimized norm-conserving Vanderbilt (ONCV) pseudopotential (Hamann, 2013) built with the Perdew–Burke–Ernzerhof (PBE) exchange correlation functional (Perdew et al., 1996) is used for the CPMD simulations. In addition, the PBEsol exchange correlation of the GGA (Perdew et al., 2008) is used to calculate static properties such as lattice constants and defect formation energies, which are then compared to the PBE results. The reciprocal space integration is carried out at the gamma point in the first Brillouin zone. For calculations at zero temperature and zero pressure, the atomic configurations and unit cell geometries were relaxed by energy minimization via the Broyden–Fletcher–Goldfarb–Shanno (BFGS) algorithm, implemented in QE.

As a validation step of the DFT calculations for garnet, a few physical properties at 300 K are calculated, namely the lattice constant, the bulk modulus, the heat capacity, and the thermal expansion coefficient. To this end, we make use of the quasi-harmonic approximation (QHA) with the finite displacement method, which has been automated in the Phonopy software (Togo, 2023). QHA is a phonon-based model for the vibrational energy of a solid that assumes harmonic interatomic potentials and, concomitantly, a dependence of the phonon dispersion relation on the volume of the unit cell. One performs a series of independent simulations at different unit cell volumes to estimate the (Gibbs) free energy, from which thermodynamic properties such as the heat capacity, the thermal expansion coefficient, and the elastic modulus are obtained at elevated, but not too high temperatures (Baroni et al., 2009). To perform the QHA calculation, the relaxed lattice structure of garnet was compressed and expanded isotropically by ±6% (which corresponds to, approximately, ±10 GPa), incrementally in steps of 1%, to calculate the phonon energies.

A comparison of the results from the QHA and DFT calculations with available experimental data (Table 1) shows that using the PBE exchange-correlation functional overestimates the lattice constant and, correspondingly, underestimates the density by ca. 4%, which is a good value for crystals. With the PBEsol functional, the deviation from the experimental densities is reduced to about 1%. For both functionals, the predicted bulk modulus and thermal expansion coefficient fall within the ranges set by the experimental data, with the bulk modulus matching the experimental data within 3%. The calculated heat capacity agrees within a 5% margin with the experimental values, which spread wider than the elastic data.

### 2.2 Ab initio molecular dynamics simulations

For the study of H diffusion in pyrope garnet, we performed simulations in the NPT ensemble via the Car–Parrinello scheme of Quantum ESPRESSO (Car and Parrinello, 1985), coupled to a Nose–Hoover thermostat (Martyna et al., 1992) and a Parrinello–Rahman barostat (Parrinello and Rahman, 1981). Protonated perfect pyrope lattices and defective lattices with either a Mg, Si, or Al vacancy were modeled at pressure–temperature conditions of 10 and 20 GPa and 2000, 2500, and 3000 K, respectively. To this end, we inserted two interstitial H atoms into the perfect lattice and, for the defective lattice, one Mg, Al, or Si atom was substituted by 2, 3, and 4 hydrogen atoms, respectively. The fictitious electron mass of the Car–Parrinello scheme was set to 100 atomic units, and the integration timestep was fixed to 0.15 fs. We tested the importance of the electron mass, which puts limits on the timestep and accuracy, by comparing the lattice constants obtained for electron masses of 50, 100, and 250 atomic units; the value 100 was found to serve as a reasonable balance. The total length of the simulated trajectories was between 150 and 250 ps. The computation time for one 200-ps long run was ca. 200 hours on the wall clock using 96 cores of type Intel Xeon Skylake 6130 in parallel.

The diffusion coefficient of hydrogen was obtained via the mean-square displacement (MSD):

$$\mathrm{MSD}(t) = \frac{1}{N}\sum_{i=1}^{N}\langle|\vec{r}_i(t+t_0) - \vec{r}_i(t_0)|^2\rangle \, ,$$

where $N$ is the number of hydrogen atoms, $\vec{r}_i(t_0)$ is the position of the $i^{\text{th}}$ H atom at the reference time $t_0$ and $\vec{r}_i(t+t_0)$ its position after the lag time $t$. The brackets $\langle\cdots\rangle$ denote an average in time, which is realized by moving $t_0$ along the simulated trajectory. In practice, we limit the lag time to 10% of the total run time to ensure small statistical noise of the MSD at long lag times. At short lag times, it holds $\mathrm{MSD}(\mathrm{t} \to 0) \simeq 3k_BT\, t^2/m_H$, which serves as a sanity check of the MSD calculation. The diffusion constant $D$ was obtained from the time derivative of the MSD at long times according to the Einstein relation (see, e.g., Höfling and Franosch (2013)): $D = \mathrm{MSD}'(t \to \infty)/6$.

## 2.3 Tracer diffusion on the potential energy surface

As an alternative to computationally expensive CPMD simulations, we simulate the tracer diffusion of H atoms (tracer MD) by constructing the potential energy surface of an interstitial hydrogen atom in a cubic pyrope unit cell, containing 160 atoms. The key approximations of tracer MD are that (i) the H atoms induce no significant change in the structure of the host lattice, (ii) they do not interact with each other (low H concentration), and (iii) the thermal fluctuations of the local environment (atomic vibrations) are neglected. Under these conditions, the atomic positions of all the atoms, except hydrogen, are fixed to the ground-state configuration of the lattice and the H atoms independently explore the highly irregular potential energy surface, which results from the forces of the lattice atoms exerted on each H atom. Such tracer MD simulations have been used in different contexts to study molecular transport in amorphous host materials (Boţan et al., 2013; Collell et al., 2015; Höfling et al., 2006; Schnyder and Horbach, 2018; Spanner et al., 2016; Voigtmann and Horbach, 2009). Tracer MD simulations are computationally relatively cheap, and the main load is on the pre-calculation of the potential energy surface. In addition to this gain in efficiency, a comparison of the tracer diffusion study with the full CPMD results allows us to assess the importance of atomic vibrations of the garnet lattice for hydrogen diffusion.

The potential energy surface was calculated *ab initio* via DFT for a relaxed pyrope unit cell (160 atoms) with one additional H atom sampling the unit cell on a cubic mesh with $50^3$ mesh, i.e., a mesh size of ca. 0.2 Å. For the potential energy surface of the perfect lattice, we took advantage of the bcc symmetry so that only 1/16 of the unit cell needed to be sampled. For the lattice with one Mg vacancy, the broken symmetry required us to calculate the energy at all nodes of the three-dimensional mesh. To obtain a sufficiently smooth potential energy surface in the entire unit cell, we used a minimal tricubic interpolation scheme (Lekien and Marsden, 2005), which, at each node, requires the value of the energy $E(x, y, z)$, the gradient ($\partial E/\partial x = -F_x$, ...), the matrix of second derivatives (e.g., $\partial^2 E/\partial x\partial y$), and the mixed third derivative, $\partial^3 E/\partial x\partial y\partial z$). For each of the 8 corners of a cubic voxel of the mesh, these 8 coefficients were calculated and stored using the ground-state energy $E$ and the force vector $(F_x, F_y, F_z)$ obtained from the DFT calculations with QE; the higher derivatives were approximated by central differences on the basis of the forces.

The stored coefficients were then used in tracer MD simulations via a custom implementation as a MATLAB script. The trajectories of an ensemble of mutually non-interacting H atoms were computed simultaneously. The H atoms were initially placed at random positions within the unit cell via a grand canonical Monte Carlo insertion scheme (Schnyder et al., 2015), with the acceptance probability $\exp(-(E(\vec{r}) - E_{min})/k_B T)$ for the potential energy $E(\vec{r})$ at the trial position $\vec{r}$, where $E_{min}$ is the minimal energy of the PES, and $k_B$ is Boltzmann's constant. Time integration was performed with the velocity Verlet algorithm coupled to a Nosé–Hoover thermostat chain of length 2 (Frenkel and Smit, 2001; Martyna et al., 1992). The time step was set to 0.15 fs, allowing us to produce 1-ns-long trajectories of 1000 hydrogen atoms in parallel, which merely took a few hours on a personal laptop.

### 2.4 MD simulations with machine-learned interatomic potentials

Despite the speed gain of tracer-MD, there is the strong assumption that all the atoms except H are immobile and that there is no interaction between the H atoms. With the goal of eliminating those assumptions and pushing the computational limitations of CPMD simulations on the duration of the generated hydrogen trajectories, we used machine learning techniques to train a deep neural network (DNN) that encodes the potential energy surface obtained from DFT calculations. Originally introduced by Behler and Parrinello, (2007), ML-based interatomic potentials today exist in a great variety of representations (Behler, 2021; Deringer et al., 2021; Unke et al., 2021). Here, we used the DeePMD-kit package (Wang et al., 2018), which implements a direct mapping between local atomic configurations and the corresponding potential energy (Zhang et al., 2018a). This type of DNN potential is similar to the second generation DNN potentials in Behler's classification; however, the demanding calculation of atomic symmetry functions is bypassed. Instead, for each atom, an embedding net is fed with a sorted collection of relative positions of neighboring atoms, which respects the translational, rotational, and permutational symmetries as well as the periodic boundary conditions. We chose the smooth edition of the two-atom angular-type embedding descriptor (DeepPot-SE, called '*se_e2_a*') with an embedding net of three layers of sizes 25, 50, and 100 and a smooth truncation at the cutoff radius 6 Å (Zhang et al., 2018b). These descriptors are then input to a feed-forward neural network of three hidden layers, each of size 360. This choice of the descriptor and of the network topology is well-suited to simulate diffusion in a crystalline compound such as garnet.

For the training of the potential energy and force calculated via DFT, we randomly selected 1000 atomic configurations from each CPMD run at different P-T conditions described in the CPMD section and performed a self-consistent field DFT calculation on each configuration. Additional training data were included from CPMD simulations on a perfect lattice (without hydrogen atoms) at lower temperatures of 300, 800, and 1800 K and pressures of 1 and 10 GPa for a shorter time span of 10 ps, with the aim of covering a wider range of P–T conditions for more accurate predictions of the lattice parameters. The root-mean-square errors of energy and force were tracked. Each model is trained after $10^6$ steps, with the learning rate decreasing from $10^{-3}$ to $10^{-7}$. The training can favor either the energy or the force. Here, the weighting factor of the energy was increased during training from 0.02 to 1, and the weight of the force was decreased from $10^3$ to 1. The resulting root mean square error of the energy is less than $10^{-3}$ eV and of the force less than 0.1 eV $Å^{-1}$; the conversion factor to the SI unit kJ $mol^{-1}$ is 1 eV = 96.485 kJ $mol^{-1}$. The energies and forces obtained from the trained DNN potential match well with the results from the DFT calculations (Figure 2). After training, we used the Large-scale Atomic/Molecular Massively Parallel Simulator (LAMMPS) (Plimpton, 1995) to carry out MD simulations with the trained interatomic potential. The MD simulations were performed with the velocity Verlet integrator coupled to a Nose–Hoover thermostat (Martyna et al., 1992) and a Parrinello–Rahman barostat (Parrinello and Rahman, 1981). The time step was set to 0.2 fs to account for the comparably fast motion of hydrogen atoms. The system size is known to affect the diffusivity in a random host structure (Höfling et al., 2008). Potential corrections of the diffusivity due to the finite system size were tested systematically by replicating the garnet unit cell with 2 interstitial H atoms into 1–1–1, 2–2–2 and 3–3–3 supercells and running MLIP-MD simulations for 1 ns. The 2-2-2 or 3-3-3 supercells were found to be sufficiently accurate.

To validate the MLIP-MD simulations and the trained DNN potential, we computed the lattice constant and the radial distribution function for a perfect pyrope lattice under high P–T conditions. Both quantities agree excellently with the reference results obtained from the CPMD simulations (Figure. 2a,b).

## 2.5 Calculation of defect formation energies

The formation energies $\Delta E$ of point defects were calculated from first principles using DFT via the balance equation (Zhang and Northrup (1991):

$$\Delta E = E_{\text{defect}} - E_{\text{perfect}} - \sum_{i=1}^{n} n_i \mu_i ,$$

where $E_{\text{defect}}$ and $E_{\text{perfect}}$ are the energies of the defective and perfect lattices, respectively, $n_i$ is the number of atoms of species $i$, and $\mu_i$ is the chemical potential per atom. Since this study is concerned with formation reactions of charge-balanced defects in an insulator, there is no contribution from the transfer of reservoir electrons. Furthermore, owing to the large unit cell of garnet, we neglect a potential defect–defect interaction energy due to the periodic boundaries. The energies $E_{\text{defect}}$ and $E_{\text{perfect}}$ were calculated by relaxing the atomic configurations via energy minimization. The formation energies thus represent the condition of zero temperature and zero pressure.

For a multi-component crystal such as pyrope garnet ($Mg_3Al_2Si_3O_{12}$), the chemical potential for each atomic species is not directly accessible (Verma and Karki, 2009). This is because the chemical potentials of Mg, Al, Si and O satisfy merely the constraint $\mu_{\mathrm{Mg_3Al_2Si_3O_{12}}} = 3\mu_{\mathrm{Mg}} + 2\mu_{\mathrm{Al}} + 3\mu_{\mathrm{Si}} + 12\mu_{\mathrm{O}}$, where the left-hand side is obtained from DFT calculations but the individual contributions on the right-hand side cannot be determined unambiguously. In this case, only the full Schottky defect can be calculated (Verma and Karki, 2009), whereas Schottky pairs such as $[\mathrm{V}''_{\mathrm{Mg}} + \mathrm{V}^{\bullet\bullet}_{\mathrm{O}}]^{\times}$ require the chemical potential of Mg and O atoms to be known separately. In Kröger–Vink notation (Kröger and Vink, 1956), $\mathrm{V}''_{\mathrm{Mg}}$ refers to a two-fold negatively charged Mg vacancy and $\mathrm{V}^{\bullet\bullet}_{\mathrm{O}}$ to a two-fold positively charged O vacancy; the superscripted × denotes the charge neutrality of the defect. To proceed, we resort to the chemical potential of periclase to obtain the reaction enthalpy of the Mg–O Schottky pair; similarly, we used quartz for $SiO_2$ and corundum for $Al_3O_2$. To reduce finite-size effects, we used a 2-by-2-by-2 supercell for the DFT calculations of periclase, quartz and corundum. As an example, the defect formation reaction:

$$\mathrm{Mg}^{\times}_{\mathrm{Mg}} + \mathrm{O}^{\times}_{\mathrm{O}} = \mathrm{V}''_{\mathrm{Mg}} + \mathrm{V}^{\bullet\bullet}_{\mathrm{O}} + \mathrm{MgO},$$

has the reaction energy:

$$\Delta E = E_{\mathrm{V}''_{\mathrm{Mg}}} + E_{\mathrm{V}^{\bullet\bullet}_{\mathrm{O}}} + E_{\mathrm{MgO}} - 2E_{\mathrm{perfect}} \,,$$

where $E_{\mathrm{V}''_{\mathrm{Mg}}}$ is the energy of the lattice with one $\mathrm{V}''_{\mathrm{Mg}}$ vacancy, $E_{\mathrm{V}^{\bullet\bullet}_{\mathrm{O}}}$ is the energy of the lattice with one $\mathrm{V}^{\bullet\bullet}_{\mathrm{O}}$ vacancy, $E_{\mathrm{MgO}}$ is the energy per formula of periclase, and $E_{\mathrm{perfect}}$ is the energy of the perfect pyrope unit cell.

We studied four types of point defects: Schottky defects, Frenkel defects, protonation defects, and antisite defects (Table 2). To test the DFT approximation error, we have again obtained results for the PBE and PBEsol functionals. In general, the energy difference between PBE and PBEsol is within 80 kJ mol$^{-1}$, except for the Si-O Schottky defect, for which the difference is ca. 120 kJ mol$^{-1}$. This is probably caused by a larger difference in the atomic position after relaxation.

For the protonation defects, we used water as the H source so that the energy of the O–H bond is included in the formation energy. Furthermore, we determined the formation energy of the protonation defects at 300 K since water does not exist in its liquid form at 0 K, where standard DFT calculations are carried out. For the chemical potential of liquid water, we followed two different routes: first, we separately calculated the energy of isolated $H_2$ and $O_2$ molecules in a large simulation box (with edge lengths of 20 Å) using DFT with energy minimization to relax the distance between the atoms. We then approximated the chemical potential of the $H_2O$ molecule by adding the formation enthalpy of liquid water, 285.83 kJ mol$^{-1}$, at 300 K (Chase, 1998). The second route consisted of directly performing a CPMD simulation with 64 water molecules at ambient conditions, 300 K and $10^5$ Pa, which directly yields the energy of liquid water. The latter value is only ca. 5 kJ mol$^{-1}$ larger than the first result, and the small difference also serves as a validation of the CPMD simulation. For the energy of the pyrope lattices with and without protonation defects, we used QHA (see above) to calculate the isobaric heat capacity $C_p(T)$ for a number of temperatures in the range from $T = 0$ K to 300 K. The lattice energies at 300 K were then obtained via numerical integrating $C_p(T)$.

## 3 Results

### 3.1 Defect formation energy

We have calculated the formation energy for a number of charge-balanced point defects in pyrope garnet (Table 2). In general, the formation energies of protonation defects are considerably lower than those of the other types of defects studied, namely Schottky, Frenkel, and antisite defects. The protonated Si vacancy $(4\mathrm{H})_{\mathrm{Si}}^{\times}$ has a formation energy that is 2–3 times lower than that of the $(2\mathrm{H})_{\mathrm{Mg}}^{\times}$ and $(3\mathrm{H})_{\mathrm{Al}}^{\times}$ protonation defects. The relaxed atomic structures of the protonation defects (Figure 3) show that for $(4\mathrm{H})_{\mathrm{Si}}^{\times}$, the four H atoms are equally separated, with each H atom attached to one O atom in the tetrahedral site. For $(2\mathrm{H})_{\mathrm{Mg}}^{\times}$, the two H atoms are bound to two O atoms on opposite sides, whereas for the relaxed $(3\mathrm{H})_{\mathrm{Al}}^{\times}$ octahedral site, one H atom sits at the lower (or upper) apex and the other two H atoms bond to the O atoms at the basal plane.

The second most energetically favored defect type are antisite defects, with the Si–Al pair having the lowest formation energy (approx. 110 kJ $mol^{-1}$). This is consistent with recent work by Subasinghe et al. (2022), which showed that Si–Al antisite defects in almandine garnet have a lower formation energy than other defects (protonation defects were not considered in that study). Among the Schottky defects, the Mg–O pair has the lowest formation energy (ca. 310 kJ $mol^{-1}$), which is well below the formation energies of the other two Schottky pairs, Al–O (ca. 540 kJ $mol^{-1}$) and Si–O (ca. 800 kJ $mol^{-1}$). The average value is reported based on PBE and PBEsol functionals. Among the investigated defects, Frenkel defects have the highest formation energies, ranging between 560 kJ $mol^{-1}$ and almost 1000 kJ $mol^{-1}$, which is likely caused by the dense atomic structure of garnet, yielding a high energetic cost for heavier interstitial atoms.

### 3.2 Potential energy surface

Pyrope garnet has a complex potential energy surface due to its large unit cell containing 160 atoms. The potential energy surface of an interstitial hydrogen atom in the perfect garnet lattice (Figure 4b–d) forms individual "pockets" for energies below 80 kJ $mol^{-1}$. These pockets are interconnected by the isosurface at 100 kJ $mol^{-1}$, which suggests that a hydrogen atom that has a kinetic energy of less than 80 kJ $mol^{-1}$ will be unable to escape the energy pocket, assuming that the other atoms in the garnet lattice are immobile. If the kinetic energy of the hydrogen atom is greater than 100~120 kJ $mol^{-1}$, it may travel freely within the garnet.

For a defective lattice with a Mg vacancy, however, a large energy sink of 300 kJ $mol^{-1}$ appears at the site of the vacancy (Figure 4f–h) and a kinetic energy of at least 300 to 450 kJ $mol^{-1}$ is required to connect the energy isosurface around the vacancy site to the boundary of the unit cell. Thus, the kinetic energy of the hydrogen atom needs to be either as high as or above 300 kJ $mol^{-1}$, or the heavier atoms (Mg/Si/Al/O) must move so that a gap opens for the H atom to escape. The latter effect of the atomic vibration of the heavier atoms on hydrogen diffusion will be studied below by means of MD simulations.

### 3.3 Temporal resolution of the hydrogen diffusion process

The diffusion of H atoms is quantified with temporal resolution by the MSD, which also yields the the hydrogen diffusion coefficient *D* at long lag times. We discuss the hydrogen MSD exemplarily for the perfect pyrope lattice and a defective lattice with a Mg vacancy, both at a temperature of 2500 K and a pressure of 10 GPa. (Figure 5a,b). We also compare results from CPMD, MLIP-MD, and tracer-MD simulations; while the former two are expected to agree, tracer-MD relies on the additional assumption that the vibrational motion of the lattice atoms has a negligible effect on H diffusion. For the perfect lattice, all three MSD curves are similar at both short and long-time scales, with the exception that the long-time data obtained from the tracer-MD simulation are suppressed by a factor of about 2. In particular, all three methods yield inertial ("ballistic") motion at short time scales (below 10 fs) with a quadratic increase in time, $\mathrm{MSD(t)} \propto t^2$. At long time scales (above ca. 1 ps), one observes the linear increase in time that is characteristic of diffusion, i.e., $\mathrm{MSD(t)} = 6Dt$. In between, at intermediate time scales, the tracer-MD results differ from those of the other two methods, which leads to the abovementioned spurious suppression of the diffusivity at long times. The MSD at these time scales corresponds to hydrogen displacements of 1 to 2 Å, which we interpret as H movements within, e.g., cation or interstitial sites. We conclude that lattice vibrations play a role in such processes and contribute to H diffusion.

For a defective lattice with a Mg vacancy, we find a reduction in the MSD compared to the perfect-lattice case and a corresponding decrease in the diffusion coefficient by a factor of about 2 (Figure 5b). In addition, the diffusive regime is entered only 100 times later, at lag times above ca. 10 ps, corresponding to hydrogen movements of the order of the lattice constant. Whereas a reduction in diffusivity is expected for the escape even from a simple potential trap, the observed long delay is typical for motion in a complex environment (Höfling and Franosch (2013), giving rise to a regime of subdiffusive motion, which has a sublinear increase of MSD, i.e. $\mathrm{MSD}(t) \propto t^{\alpha}$ with exponent $\alpha < 1$; the fitted value is $\alpha \approx 0.6$. We note that the CPMD and MLIP-MD results for the MSD agree well, with the exception of lag times of approx. 100 fs; in particular, they yield similar diffusion coefficients. However, MLIP-MD allows us to access much longer lag times and yields data with a much better signal-to-noise ratio than CPMD does. The MSD of the tracer MD simulations coincides with the CMPD and MLIP-MD results in the inertial regime (short time scales). However, it deviates significantly for times longer than 30 fs and shows pronounced subdiffusive behavior up to about 20 ps, with a small fitted exponent of $\alpha \approx 0.2{\sim}0.3$. Such small subdiffusion exponents are known from diffusion in strongly heterogeneous, arrested random media (Höfling et al., 2006; Spanner et al., 2016). The displacement at the upper end of this subdiffusive window is merely 1.5 Å, and we conclude that most of the H atoms would still be trapped at the vacancy site after 20 ps if the vibrations of the lattice atoms are switched off as in tracer MD. We have not followed the motion for longer time scales, yet we anticipate that the crossover to long-time diffusion will occur at much longer times, where the MSD becomes comparable to the lattice constant squared.

Conversely, the H atoms escape the vacancy site more easily for a more realistic lattice dynamics, as obtained from CPMD and MLIP-MD simulations; the process is depicted in Figure 6: Initially, the H atom is located in a cage of eight O atoms. Then, it moves toward an O atom at one of the corners of the vibrating O cage. Owing to the vibrations, a gap opens and the H atom escapes to the outside of the O cage. The escape occurs within less than a few femtoseconds in a much longer interval (several picoseconds to several 100 ps), during which the H and O atoms vibrate around their energy-relaxed positions.

For perfect pyrope garnet, we have calculated the hydrogen diffusion coefficients for temperatures $T$ ranging between 1500 K and 3000 K, which display Arrhenius behavior, $D(T) \propto \exp(-Q/k_B T)$ with the activation energy $Q$ (Figure 5c). The three simulation approaches yield consistent results, as expected from the above. For the lowest temperatures, however, the diffusive regime in time could only be reached within the tracer-MD simulations with the caveat that these results provide only a lower bound on the diffusivity because the vibrations of the garnet lattice are not included as discussed above. The activation energy was determined from the tracer-MD data to be ca. 93 kJ mol$^{-1}$; from the MLIP-MD data, we obtain ca. 73 kJ mol$^{-1}$.

### 3.4 Dependence of the hydrogen diffusion coefficient on the concentration of vacancies and hydrogen

Lattice defects, such as vacancies, play a crucial role in hydrogen diffusion and may also influence the activation energy. In addition, both quantities are anticipated to depend on the concentration of H atoms. We explored these effects by systematically varying the number $N_{\text{vac}}$ of $V''_{\text{Mg}}$ and $V''_{\text{Si}}$ vacancies and the number $N_{\text{H}}$ of H atoms in MLIP-MD simulations of 3-by-3-by-3 garnet supercells, which have a volume of ca. 4.3 nm$^3$; specifically, we varied $N_{\text{H}}$ between 8 and 48 and $N_{\text{vac}}$ between 4 and 16 and considered pyrope garnet at two temperatures, 2500 K and 3000 K, under 10 GPa pressure conditions (Figure 7). The results show large variations in both the diffusion coefficient (by factors of 4 for Mg and 12 for Si), and the activation energy (by factors of 3–4). For both Mg and Si vacancies, upon increasing the number of vacancies, the H diffusivity decreases. Upon increasing the number of H atoms, the H diffusivity increases. The activation energy shows the opposite trend: it increases upon increasing the number of Mg or Si vacancies, and it decreases with increasing hydrogen concentration.

For fixed ratios $N_{\text{H}}/N_{\text{vac}}$ (i.e., along straight lines in Figure 7), the obtained diffusion coefficients vary only weakly, and they increase as the ratio $N_{\text{H}}/N_{\text{vac}}$ increases from 1 to 4 and beyond. Considering charge balance, one may expect that the increase in diffusivity upon increasing $N_{\text{H}}/N_{\text{vac}}$ would stop at the ratio of two H atoms per vacancy. However, this expectation is not supported by our data, which suggests that more than two H atoms are required to saturate one Mg vacancy site and that many H atoms are sitting at different interstitial sites. The same argument applies to Si vacancies.

Concerning the activation energies $Q$, the increase of $Q$ with the number of vacancy sites is faster for Si than for Mg vacancies. For example, in a 3-by-3-by-3 supercell with 16 Si vacancies and 12 or 24 H atoms, the activation energy reaches 220 to 230 kJ mol$^{-1}$, which is about twice as high as the value of 100 to 160 kJ mol$^{-1}$ obtained for Mg vacancies. We conclude that, overall, it is more difficult for H atoms to escape a Si vacancy than a Mg vacancy.

On the basis of the obtained H diffusion coefficients and activation energies as functions of the H/vacancy ratio, we classify H diffusion into two main regimes: a diffuser-dominated regime and a vacancy-dominated one. The boundary between these two regimes is not sharply defined, but is roughly located near $N_{\text{H}}/N_{\text{vac}} \approx 2$ for the Mg- and Si-type vacancies (Figure 7). The diffuser-dominated regime, where $N_{\text{H}}/N_{\text{vac}}$ is large, is characterized by a high diffusivity of H atoms and a low activation energy, which is close to the value of the perfect lattice. On the other hand, the vacancy-dominated regime, where $N_{\mathrm{H}}/N_{\mathrm{vac}}$ is small, exhibits a low H diffusivity and a high activation energy.

# 4 Discussion

## 4.1 Hydrogen diffusion kinetics

Based on the calculation of the defect formation energy (Table 2), the Si protonation defect $(4\mathrm{H})_{\mathrm{Si}}^{\times}$ is energetically most favorable. Quartz and $H_2O$ buffers are common in crustal conditions. The formation of $(4\mathrm{H})_{\mathrm{Si}}^{\times}$ might be limited by the transition state during the removal of Si from the tetrahedral site, whose activation energy is relatively high, ca. 345±56 kJ mol$^{-1}$ from experimental constraints (Shimojuku et al., 2014). The second most energetically favored defect, Mg protonation $(2\mathrm{H})_{\mathrm{Mg}}^{\times}$, can be formed more easily because the diffusion of Mg cations has a lower energy barrier of approximately 258.0±7.4 kJ mol$^{-1}$ (Chu and Ague, 2015) and 244.2±5.3 kJ mol$^{-1}$ (Carlson, 2006). On the basis of the present MD simulation results, the transport of H atoms in garnet is not governed by defect migration in the investigated temperature range.

Our simulation results for H diffusion suggest that hydrogen can easily migrate in a perfect pyrope lattice with a low energy barrier of less than 100 kJ mol$^{-1}$. The activation energy is a function of the relative abundance of vacancies (e.g., $[\mathrm{V}_{\mathrm{Mg}}'']$) and the hydrogen concentration. We have identified two main regimes, namely diffuser-dominated and vacancy-dominated (Figure 7). In the former, H diffusers are abundant and the vacancies are filled by H atoms, which results in an activation energy close to the value of the perfect garnet lattice. In the vacancy-dominated regime, few H atoms face a large number of "traps" and their number is not sufficient to fill all vacancy sites. This leads to an increase of the activation energy as a function of the vacancy concentration.

The CPMD and MLIP-MD simulations revealed that H atoms may escape the trapping site and become interstitial before they find another energetically favored location (Figure 6). During this process, the vibrations of heavier atoms such as O, Mg, Al, and Si play a vital role, which is corroborated by the observation that the H atoms do not escape the vacancy site, even not after long times of 100 ps, if the lattice vibrations are artificially switched off as in the tracer MD simulations. Interestingly, the H atoms perform a subdiffusive motion over about three decades in temporal resolution (from ca. 30 ns to ca. 20 ps), with exponents (i) $\alpha \approx 0.6$ for the full lattice dynamics (MLIP-MD data in Figure 5b) and (ii) $\alpha \approx 0.2{\sim}0.3$ without lattice vibrations (tracer MD data in Figure 5b). These findings resemble observations in basic models of (i) arrested liquids after vitrification (Voigtmann and Horbach, 2009) and (ii) amorphous solids that feature long trapping times (Höfling et al., 2006; Spanner et al., 2016), respectively. The occurrence of subdiffusion implies motion in a complex environment, either due to a spectrum of energy scales or a spectrum of length scales or both (Höfling and Franosch (2013). This suggests that the H atoms move in a highly irregular potential energy surface and that the H escape from a vacancy site is not well pictured by barrier crossing from a simple energy well.

## 4.2 Garnet composition effect

Solid solutions of natural garnets contain various endmembers at large mantle depths, such as almandine (Fe), grossular (Ca), spessartine (Mn), and andradite ($Ca_3Fe_2(SiO_4)_3$), along with majorite ($Mg_3(MgSi)(SiO_4)_3$). While crustal garnets are mainly almandine-rich, mantle garnets are dominated by pyrope-rich compositions. Both main cations on the eightfold-coordinated site, $Fe^{2+}$ and $Mg^{2+}$, have similar ionic radii (Armbruster et al., 1992), which implies that their relative diffusivities and their effects on hydrogen diffusion should be very similar (Chu and Ague, 2015). We therefore tested, in addition to pyrope, the effect of the grossular ($Ca_3Al_2(SiO_4)_3$) component on the diffusion coefficient. Because Ca has a much larger ionic radius than Mg, Fe, and Mn do, the effect of Ca on the diffusion coefficient is expected to be the largest. We have investigated this for the perfect grossular lattice by employing the tracer MD diffusion model, following the same procedure used for pyrope. The obtained H diffusion coefficient in grossular garnet is lower, by a factor of less than two, compared to that of pyrope garnet within the same tracer MD model (Figure 8). The activation energy of grossular (105.9 kJ mol$^{-1}$) was found to be slightly greater than that of pyrope (93.3 kJ mol$^{-1}$) by ca. 12 kJ mol$^{-1}$. A possible explanation for these differences is in the larger ionic radius of Ca in grossular. Beyond that, our results do not indicate a substantial effect of the composition of Ca and Mg on H diffusivity. While we have tested the composition effect only for perfect lattices, we anticipate that the H diffusivity in defective lattices with either Ca or Mg vacancies differ largely. Notably, our modeled results are in agreement with the experimental evidence that H in grossular has a relatively high activation energy and a low diffusion coefficient (Kurka et al., 2005; Wang et al., 1996).

### 4.3 Advantages and disadvantages of the employed simulation approaches

Among the three simulations approaches we used here, CPMD makes the mildest approximations to the physics of the problem. Hence, it is considered the most accurate method for studying hydrogen diffusion and it serves as a reference for the other approaches. In practice, the application of CPMD is limited by its high computational demand, in particular, if large supercells shall be modeled or if long trajectories are needed to follow sluggish diffusion processes.

We have demonstrated here that MLIP-MD is a useful alternative to CPMD because it allows one to expand the model size or perform longer simulations. Compared with CPMD, the trained MLIP-MD generally performs well in correctly reproducing the lattice constant, radial distribution function and diffusivity under high P–T conditions. In this study, we tested only one type of MLIP, the deep neutral network (DNN) potential (Zhang et al., 2018a) implemented in the DeePMD-kit software (Wang et al., 2018). Currently, numerous machine learning approaches for realistic interaction potentials are developed in parallel and are employed in various fields; further notable approaches are Behler–Parrinello neural network potentials (Behler and Parrinello, 2007; Behler, 2021) and Gaussian approximation potentials (Bartõk and Csányi, 2015; Deringer et al., 2021). Some of these methods have been successfully applied to geomaterials under high P–T conditions, such as $SiO_2$ polymorphs (Erhard et al., 2022). Recently, the ReaxFF force field for earth minerals has also been established for a limited range of elements (Si, Al, O, H, Na, K) (Zhang et al., 2024). Although not suitable for garnets in the current state, the reactive force field may also play a vital role in modeling petrological processes and obtaining petrophysical mineral properties in the future. There is no simple answer to the question of which is the “best” machine learning force field or empirical force field to choose, even after a systematic study (Pinheiro et al., 2021), and this topic is beyond the scope of this work. A fundamental limitation of MLIP approaches, which is current consensus, is that the conditions (such as P–T) and the modelled structures of the training data should fully cover all possible situations of the actual simulations. For validation, the results, including physical observables such as the lattice constant, radial distribution function, MSD, or velocity autocorrelation function, need to be carefully examined and compared with those of reference methods, e.g., CPMD. However, care still needs to be taken in handling complicated transport and vibrational behavior. For example, in our study, although MLIP-MD captures well the kinetics of H diffusion at time scales longer than 10 ps and the inertial motion at short time scales (less than 10 ns), it yields some deviations from the CPMD reference for the MSD at the intermediate time scales. However, our tests suggest that this issue has only a minor effect on the MLIP-MD results for the H diffusion coefficients.

Finally, tracer MD simulations follow only the motion of H atoms and treat the lattice atoms as fixed in space, thereby switching off lattice vibrations. Despite this oversimplification, we have found that the approach is capable of simulating hydrogen diffusion in perfect lattices, where energy barriers are low. The advantage is that one can obtain good estimates of the hydrogen diffusion coefficient at very low temperatures, even at temperatures as low as 1500 K, using a personal laptop (Figure 5). This is extremely difficult to achieve even for MLIP-MD. In the presence of a potential energy well, e.g., at the Mg vacancy site, we have shown that tracer MD yields qualitatively very different results than those of CPMD and MLIP-MD. However, this “failure” is also informative and highlights the importance of the vibrations of the heavier lattice atoms (Al, Mg, Si, O) for the migration of the much lighter H atoms. We speculate further that vibrations may be more important for H diffusion than quantum mechanical tunneling, which is resolved by neither of the methods used here.

### 4.4 Comparison with experiments and petrological implications

Hydrogen diffusion in garnet has been studied experimentally using deuteration (D-H replacement) and deprotonation techniques. Figure 8 shows experimental results on the diffusion coefficient from works by Blanchard and Ingrin (2004); Kurka et al., (2005); Phichaikamjornwut et al. (2012); Reynes et al., (2018); Zhang et al. (2015) and Wang et al. (1996) together with our simulated diffusion coefficients. Among the experimental data, the diffusion coefficient of pyrope by Zhang et al. (2015) is the highest, whereas the other sets of data cover more than three orders of magnitude for temperatures ranging between 700 °C and 1000 °C. The experimental activation energies overall lie between 100 and 350 kJ $mol^{-1}$, where the value obtained from a D-H replacement experiment by Zhang et al. (2015) for andradite is the lowest (ca. 96 kJ $mol^{-1}$) and Kurka et al. (2005) report the second lowest value (ca. 102 kJ $mol^{-1}$). The deprotonation experiment by Wang et al. (1996) yields a higher activation energy of ca. 254 kJ $mol^{-1}$, and Kurka et al. (2005) found ca. 323 kJ $mol^{-1}$. Our simulated activation energies for perfect pyrope (93 kJ $mol^{-1}$) and perfect grossular (106 kJ $mol^{-1}$) are consistent with the D-H replacement experiments by Zhang et al. (2015); Blanchard and Ingrin (2004) and Kurka et al. (2005). The mismatch with the deprotonation experiments may be caused by the systematic introduction of vacancies, driving the diffusion mechanism from the hydrogen-dominated regime to the vacancy-dominated regime (see the gray shaded area in Figure 8).

The present AIMD simulation results can provide an explanation for the paradox in the deprotonation experiment by Wang et al. (1996). As experimental heating time increases, the apparent H diffusion coefficient decreases, i.e., less remaining hydrogen leads to slower diffusion. Therefore, Wang et al. (1996) interpreted their data in terms of an H-concentration dependent diffusivity, which was questioned by Kurka et al. (2005), who argued that the effect may be caused by complex FTIR spectra associated with different OH sites. Our present findings from AIMD simulations support that, as H gradually diffuses out of the garnet sample during heating, the decreasing ratio between the number of H atoms and cation vacancies leads to a greater likelihood of the vacancies to temporarily trap H atoms, thereby slowing down the H diffusion process (Figure 7). This mechanism may explain the apparent time (or H concentration) dependence of H diffusivity during the deprotonation experiment.

It was proposed that coupling between H diffusivity and H content is caused by the imbalance in the diffusivity between, e.g., the slower Si vacancy or metal vacancy and faster H atoms (Shimojuku et al., 2014). On the basis of our simulation, we conclude that H migration is not fully coupled to vacancy migration itself: H atoms can be temporarily trapped, but they can escape the vacancy site to be interstitial already after a few to a few tens of picoseconds (Figure 6). A higher vacancy concentration leads to a higher likelihood of trapping, and thus a higher activation energy (Figure 7). This explains why, in general, deprotonation experiments yield a higher activation energy (150~350 kJ $mol^{-1}$) than D-H replacement experiments do (100~140 kJ $mol^{-1}$) (Figure 8b). An exception to these ranges was reported by Reynes et al. (2018), who obtained a "bulk" activation energy of 158 kJ $mol^{-1}$ from experimental data under very high oxygen fugacity. However, even this value is still higher than the highest activation energy from the collected D-H exchange experiments. The activation energy predicted in this work from MLIP-MD simulations for the perfect lattice agrees with the values from D-H replacement experiment. These experiments probe the H-diffuser dominated regime (Figure 7); unlike as in deprotonation experiments, vacancies are not necessarily created during the D-H exchange. On the other hand, the extraction of H atoms during deprotonation experiments leaves more and more deprotonated vacancies behind, which creates an apparent time (or concentration)-dependent diffusivity. In this case, the vacancy-dominated regime prevails, which is line with the higher activation energies observed in deprotonation experiments.

The defect concentration may also vary depending on the compositional mixing, the synthesis method, the quenching rate, and the redox buffer used in the experiments (Demouchy, 2021; Mori et al., 1962). These factors could lead to a change in the hydrogen/vacancy ratio, which may explain the variability of the activation energy in experimental data from different sources obtained with the same method. Further modeling and simulation studies are needed to decipher the effects of all these factors. According to our present results, changes of the garnet composition alone, without introducing additional types of vacancies, may affect the H diffusion coefficient and activation energy by merely up to 10% (Figure 8).

In this study, we have focused on vacancies in simulation models of garnet. A real crystal may contain line defects such as dislocations or plane defects such as stacking faults. Three-dimensional volume defects, such as cracks, inclusions or pores, may also play a role in the hydrogen diffusion process. The investigation of such spatially extended defects is, in principle, possible within the kind of MLIP-MD simulations employed here due to the computational speed-up over CPMD simulations. However, such investigations will require the use of considerably larger supercells and thus substantially larger computational resources. The present results support that already point defects play a major role in slowing the kinetics of H diffusion, which we explain with temporary trapping in the potential energy wells they create. Future work may also employ MLIP-MD simulations to study the effects of other types of defects on diffusion or deformation processes in Earth's materials.

## 5 Summary

We have studied hydrogen diffusion in pyrope garnet via atomistic simulations. The limitations of the conventional CPMD approach due to its immense computational cost have been overcome by training a deep neural network potential for use in MLIP-MD simulations. Thereby, we could resolve also the long-time scales of slow diffusion processes caused by the presence of a complex energy landscape with high barriers.

Our results corroborate a high mobility of hydrogen in defect free garnet lattices, whereas hydrogen diffusivity is strongly reduced in defective lattices. In the former case, the modeled activation energies (ca. 93 kJ $mol^{-1}$ for perfect pyrope and 105 kJ/mol for perfect grossular garnet) are at the lower end of the experimentally determined range (100 to 350 kJ $mol^{-1}$), which hints at the role of defects. For its systematic investigation, we compared the formation energies of a number of point and pair defects, obtained via *ab initio* DFT calculations (Table 2), and inferred that protonation effects are more likely to appear than antisite, Schottky, or Frenkel pair defects. In particular, the protonation defects $(4H)_{Si}^{\times}$ and $(2H)_{Mg}^{\times}$ are favored due to their low formation energies of 14.6~65.5 kJ $mol^{-1}$ and 111.4~137.6 kJ $mol^{-1}$, respectively. Therefore, we have focused this study of hydrogen diffusion on Mg and Si vacancies.

Upon gradually varying the concentrations of hydrogen and cation vacancies, two regimes of hydrogen diffusion can be identified in the simulation data, depending on the ratio of H atoms per vacancy (Figure 7): a diffuser-dominated regime at high hydrogen content displays low activation energies, which we attribute to a saturation of the vacancies by hydrogen. Second, a vacancy-dominated regime at low hydrogen content, where activation energies are high and individual H atoms are temporarily trapped at the vacancy sites. This sensitivity on the hydrogen to vacancy ratio can explain the higher activation energies and their apparent time dependence observed in deprotonation experiments, where the newly formed vacancies during deprotonation drives the system towards the vacancy-dominated regime.

For a defective lattice with a Mg vacancy site, we have elucidated the complexity of the H diffusion mechanism as a multi-step process of H atoms escaping from the vacancy site (Figure 6), which appears as a well-developed subdiffusive motion at temporal resolutions of up to 10~20 ps (Figure 5b). Moreover, the comparison with results from tracer MD simulations highlights the importance of lattice vibrations on H diffusion. This efficient kind of simulation focuses on the motion of H tracers by switching off the dynamics of the heavier atoms (e.g., Si, Al, Mg, O).

Eventually, an initial study on grossular provided evidence that changes of the garnet composition alone modify the hydrogen diffusion coefficient and activation energy only mildly. Whereas the present work is concerned with vacancy defects, MLIP-MD simulations are generally suited to study hydrogen diffusion also in materials with extended defects such as dislocations and stacking faults.

**Acknowledgments**

We gratefully acknowledge the use of high-performance computing resources provided on the CURTA cluster at Freie Universität Berlin and on the supercomputer Lise of Germany's NHR Alliance at Zuse Institute Berlin. Deutsche Forschungsgemeinschaft (DFG, German Research Foundation) funded this research through the grant CRC 1114 Scaling Cascades in Complex Systems, Project No. 235221301 (subproject C09) and through Project No. 523950429.

**Data Availability Statement**

Data associated to this work including the LAMMPS script and trained MLIP data can be found in the OSF platform (https://osf.io/ayz2u/).

**References**

Armbruster, T., Geiger, C.A., Lager, G.A., 1992. Single-crystal X-ray structure study of synthetic pyrope almandine garnets at 100 and 293 K. Am. Mineral. 77, 512–521.

Baroni, S., Giannozzi, P., Isaev, E., 2009. Thermal Properties of Materials from Ab Initio Quasi-Harmonic Phonons. Rev. Mineral. Geochemistry 71, 728–744.

https://doi.org/10.1007/978-1-4899-2895-5_21

Bartõk, A.P., Csányi, G., 2015. Gaussian approximation potentials: A brief tutorial introduction. Int. J. Quantum Chem. 115, 1051–1057. https://doi.org/10.1002/qua.24927

Behler, J., 2021. Four Generations of High-Dimensional Neural Network Potentials. Chem. Rev. 121, 10037–10072. https://doi.org/10.1021/acs.chemrev.0c00868

Behler, J., Parrinello, M., 2007. Generalized neural-network representation of high-dimensional potential-energy surfaces. Phys. Rev. Lett. 98, 1–4. https://doi.org/10.1103/PhysRevLett.98.146401

Blanchard, M., Ingrin, J., 2004. Kinetics of deuteration in pyrope. Eur. J. Mineral. 16, 567–576. https://doi.org/10.1127/0935-1221/2004/0016-0567

Bosenick, A., Geiger, C.A., 1997. Powder X ray diffraction study of synthetic pyrope-grossular garnets between 20 and 295 K. J. Geophys. Res. 102, 22649–22657.

Bosenick, A., Geiger, C.A., Cemič, L., 1996. Heat capacity measurements of synthetic pyrope-grossular garnets between 320 and 1000 K by differential scanning calorimetry. Geochim. Cosmochim. Acta 60, 3215–3227. https://doi.org/10.1016/0016-7037(96)00150-0

Boţan, A., Vermorel, R., Ulm, F.J., Pellenq, R.J.M., 2013. Molecular simulations of supercritical fluid permeation through disordered microporous carbons. Langmuir 29, 9985–9990. https://doi.org/10.1021/la402087r

Car, R., Parrinello, M., 1985. Unified Approach for Molecular Dynamics and Density-Functional Theory R. Phys. Rev. Lett. 55, 2471–2474. https://doi.org/10.1378/chest.93.6.1314a

Caracas, R., Panero, W.R., 2017. Hydrogen mobility in transition zone silicates. Prog. Earth Planet. Sci. 4. https://doi.org/10.1186/s40645-017-0119-8

Carlson, W.D., 2006. Rates of Fe, Mg, Mn, and Ca diffusion in garnet. Am. Mineral. 91, 1–11. https://doi.org/10.2138/am.2006.2043

Chase, M.W., 1998. NIST-JANAF Thermochemical Tables, 4th Edition. J. Phys. Chem. Ref. Data 9, 1–1951.

Chiama, K., Gabor, M., Lupini, I., Rutledge, R., Nord, J.A., Zhang, S., Boujibar, A., Bullock, E.S., Walter, M.J., Lehnert, K., Spear, F., Morrison, S.M., Hazen, R.M., 2023. The secret life of garnets: a comprehensive, standardized dataset of garnet geochemical analyses integrating localities and petrogenesis. Earth Syst. Sci. Data 15, 4235–4259. https://doi.org/10.5194/essd-15-4235-2023

Chu, X., Ague, J.J., 2015. Analysis of experimental data on divalent cation diffusion kinetics in aluminosilicate garnets with application to timescales of peak Barrovian metamorphism, Scotland. Contrib. Mineral. Petrol. 170, 1–27. https://doi.org/10.1007/s00410-015-1175-y

Collell, J., Galliero, G., Vermorel, R., Ungerer, P., Yiannourakou, M., Montel, F., Pujol, M., 2015. Transport of Multicomponent Hydrocarbon Mixtures in Shale Organic Matter by Molecular Simulations. J. Phys. Chem. C 119, 22587–22595. https://doi.org/10.1021/acs.jpcc.5b07242

Dai, L., Karato, S. ichiro, 2009. Electrical conductivity of pyrope-rich garnet at high temperature and high pressure. Phys. Earth Planet. Inter. 176, 83–88. https://doi.org/10.1016/j.pepi.2009.04.002

Demouchy, S., 2021. Defects in olivine. Eur. J. Mineral. 33, 249–282. https://doi.org/10.5194/ejm-33-249-2021

Deringer, V.L., Bartók, A.P., Bernstein, N., Wilkins, D.M., Ceriotti, M., Csányi, G., 2021. Gaussian Process Regression for Materials and Molecules. Chem. Rev. 121, 10073–10141. https://doi.org/10.1021/acs.chemrev.1c00022

Du, W., Clark, S.M., Walker, D., 2015. Thermo-compression of pyrope-grossular garnet solid solutions: Non-linear compositional dependence. Am. Mineral. 100, 215–222. https://doi.org/10.2138/am-2015-4752

Erhard, L.C., Rohrer, J., Albe, K., Deringer, V.L., 2022. A machine-learned interatomic potential for silica and its relation to empirical models. npj Comput. Mater. 8, 1–12. https://doi.org/10.1038/s41524-022-00768-w

Farber, K., Caddick, M.J., John, T., 2014. Controls on solid-phase inclusion during porphyroblast growth: insights from the Barrovian sequence (Scottish Dalradian). Contrib. Mineral. Petrol. 168, 1089. https://doi.org/10.1007/s00410-

014-1089-0

Frenkel, D., Smit, B., 2001. Understanding molecular simulation: from algorithms to applications. Elsevier.

Giannozzi, P., Andreussi, O., Brumme, T., Bunau, O., Nardelli, M.B., Calandra, M., Car, R., Cavazzoni, C., Ceresoli, D., Cococcioni, M., others, 2017. Advanced capabilities for materials modelling with Quantum ESPRESSO. (arXiv:1709.10010v1 [cond-mat.mtrl-sci]). J. Phys. Condens. Matter 29, 465901.

Giannozzi, P., Baroni, S., Bonini, N., Calandra, M., Car, R., Cavazzoni, C., Ceresoli, D., Chiarotti, G.L., Cococcioni, M., Dabo, I., Dal Corso, A., De Gironcoli, S., Fabris, S., Fratesi, G., Gebauer, R., Gerstmann, U., Gougoussis, C., Kokalj, A., Lazzeri, M., Martin-Samos, L., Marzari, N., Mauri, F., Mazzarello, R., Paolini, S., Pasquarello, A., Paulatto, L., Sbraccia, C., Scandolo, S., Sclauzero, G., Seitsonen, A.P., Smogunov, A., Umari, P., Wentzcovitch, R.M., 2009. QUANTUM ESPRESSO: A modular and open-source software project for quantum simulations of materials. J. Phys. Condens. Matter 21. https://doi.org/10.1088/0953-8984/21/39/395502

Hamann, D.R., 2013. Optimized norm-conserving Vanderbilt pseudopotentials. Phys. Rev. B - Condens. Matter Mater. Phys. 88, 1–10. https://doi.org/10.1103/PhysRevB.88.085117

Hartwig, J., Galkin, V., 2021. Heat capacity, thermal expansion, and elastic parameters of pyrope. J. Therm. Anal. Calorim. 144, 71–79. https://doi.org/10.1007/s10973-020-09396-2

Höfling, F., Franosch, T., 2013. Anomalous transport in the crowded world of biological cells. Reports Prog. Phys. 76. https://doi.org/10.1088/0034-4885/76/4/046602

Höfling, F., Franosch, T., Frey, E., 2006. Localization transition of the three-dimensional lorentz model and continuum percolation. Phys. Rev. Lett. 96, 1–4. https://doi.org/10.1103/PhysRevLett.96.165901

Höfling, F., Munk, T., Frey, E., Franosch, T., 2008. Critical dynamics of ballistic and Brownian particles in a heterogeneous environment. J. Chem. Phys. 128.

https://doi.org/10.1063/1.2901170

Jacobsen, S.D., 2018. Effect of water on the equation of state of nominally anhydrous minerals. Water Nominally Anhydrous Miner. 321–342. https://doi.org/10.2138/rmg.2006.62.14

Jamtveit, B., Austrheim, H., Putnis, A., 2016. Disequilibrium metamorphism of stressed lithosphere. Earth Sci. Rev. 154, 1–13. https://doi.org/10.1016/j.earscirev.2015.12.002

Jamtveit, B., Hervig, R.L., 1994. Constraints on Transport and Kinetics in Hydrothermal Systems from Zoned Garnet Crystals. Science 263, 505–508.

John, T., Gussone, N., Podladchikov, Y.Y., Bebout, G.E., Dohmen, R., Halama, R., Klemd, R., Magna, T., Seitz, H.M., 2012. Volcanic arcs fed by rapid pulsed fluid flow through subducting slabs. Nat. Geosci. 5, 489–492. https://doi.org/10.1038/ngeo1482

Kaatz, L., Reynes, J., Hermann, J., John, T., 2022. How fluid infiltrates dry crustal rocks during progressive eclogitization and shear zone formation: insights from H2O contents in nominally anhydrous minerals. Contrib. Mineral. Petrol. 177, 72. https://doi.org/10.1007/s00410-022-01938-1

Kohn, W., Sham, L.J., 1965. Self-Consistent Equations Including Exchange and Correlation Effects. Phys. Rev. 140, A1133–A1138.

Kröger, F.A., Vink, H.J., 1956. Relations between the Concentrations of Imperfections in Crystalline Solids. Solid State Phys. - Adv. Res. Appl. 3, 307–435. https://doi.org/10.1016/S0081-1947(08)60135-6

Kühne, T.D., 2014. Second generation Car-Parrinello molecular dynamics. Wiley Interdiscip. Rev. Comput. Mol. Sci. 4, 391–406. https://doi.org/10.1002/wcms.1176

Kurka, A., Blanchard, M., Ingrin, J., 2005. Kinetics of hydrogen extraction and deuteration in grossular. Mineral. Mag. 69, 359–371. https://doi.org/10.1180/0026461056930257

Lager, G.A., Armbruster, T., Rotella, F.J., Rossman, G.R., 1989. OH substitution in garnets: X-ray and neutron diffraction, infrared, and geometric-modeling studies.

Am. Mineral. 74, 840–851.

Lekien, F., Marsden, J., 2005. Tricubic interpolation in three dimensions. Int. J. Numer. Methods Eng. 63, 455–471. https://doi.org/10.1002/nme.1296

Luo, H., Karki, B.B., Ghosh, D.B., Bao, H., 2021. Deep neural network potentials for diffusional lithium isotope fractionation in silicate melts. Geochim. Cosmochim. Acta 303, 38–50. https://doi.org/10.1016/j.gca.2021.03.031

Martyna, G.J., Klein, M.L., Tuckerman, M., 1992. Nose Hoover chains: The canonical ensemble via continuous dynamics. J. Chem. Phys. 2635–2644.

Marx, D., Hutter, J., 2010. Ab Initio Molecular Dynamics Basic Theory and Advanced Methods.

Milani, S., Nestola, F., Alvaro, M., Pasqual, D., Mazzucchelli, M.L., Domeneghetti, M.C., Geiger, C.A., 2015. Diamond-garnet geobarometry: The role of garnet compressibility and expansivity. Lithos 227, 140–147. https://doi.org/10.1016/j.lithos.2015.03.017

Mori, T., Meshii, M., Kauffman, J.W., 1962. Quenching rate and quenched-in lattice vacancies in gold. J. Appl. Phys. 33, 2776–2780. https://doi.org/10.1063/1.1702548

Padrón-Navarta, J.A., Hermann, J., O'Neill, H.S.C., 2014. Site-specific hydrogen diffusion rates in forsterite. Earth Planet. Sci. Lett. 392, 100–112. https://doi.org/10.1016/j.epsl.2014.01.055

Parrinello, M., Rahman, A., 1981. Polymorphic transitions in single crystals: A new molecular dynamics method. J. Appl. Phys. 52, 7182–7190. https://doi.org/10.1063/1.328693

Peng, Y., Deng, J., 2024. Hydrogen Diffusion in the Lower Mantle Revealed by Machine Learning Potentials. J. Geophys. Res. Solid Earth 129. https://doi.org/10.1029/2023JB028333

Perdew, J.P., Burke, K., Ernzerhof, M., 1996. Generalized gradient approximation made simple. Phys. Rev. Lett. 77, 3865–3868. https://doi.org/10.1103/PhysRevLett.77.3865

Perdew, J.P., Ruzsinszky, A., Csonka, G.I., Vydrov, O.A., Scuseria, G.E., Constantin,

L.A., Zhou, X., Burke, K., 2008. Restoring the density-gradient expansion for exchange in solids and surfaces. Phys. Rev. Lett. 100, 1–4. https://doi.org/10.1103/PhysRevLett.100.136406

Phichaikamjornwut, B., Skogby, H., Ounchanum, P., Limtrakun, P., Boonsoong, A., 2012. Hydrous components of grossular-andradite garnets from Thailand: thermal stability and exchange kinetics. Eur. J. Mineral. 24, 107–121. https://doi.org/10.1127/0935-1221/2011/0023-2146

Pinheiro, M., Ge, F., Ferré, N., Dral, P.O., Barbatti, M., 2021. Choosing the right molecular machine learning potential. Chem. Sci. 12, 14396–14413. https://doi.org/10.1039/d1sc03564a

Plimpton, S., 1995. Fast parallel algorithms for short-range molecular dynamics. J. Comput. Phys. https://doi.org/10.1039/c7sm02429k

Plümper, O., John, T., Podladchikov, Y.Y., Vrijmoed, J.C., Scambelluri, M., 2017. Fluid escape from subduction zones controlled by channel-forming reactive porosity. Nat. Geosci. 10, 150–156. https://doi.org/10.1038/ngeo2865

Putnis, A., John, T., 2010. Replacement processes in the earth’s crust. Elements 6, 159–164. https://doi.org/10.2113/gselements.6.3.159

Reynes, J., Jollands, M., Hermann, J., Ireland, T., 2018. Experimental constraints on hydrogen diffusion in garnet. Contrib. Mineral. Petrol. 173, 1–23. https://doi.org/10.1007/s00410-018-1492-z

Rossman, G., Aines, R.D., 1991. The hydrous components in garnets: Grossular-hydrogrossular. Am. Mineral. 76, 1153–1164.

Schnyder, S.K., Horbach, J., 2018. Crowding of Interacting Fluid Particles in Porous Media through Molecular Dynamics: Breakdown of Universality for Soft Interactions. Phys. Rev. Lett. 120, 78001. https://doi.org/10.1103/PhysRevLett.120.078001

Schnyder, S.K., Spanner, M., Höfling, F., Franosch, T., Horbach, J., 2015. Rounding of the localization transition in model porous media. Soft Matter 11, 701–711. https://doi.org/10.1039/c4sm02334j

Shimojo, F., Hoshino, K., 2001. Microscopic mechanism of proton conduction in perovskite oxides from ab initio molecular dynamics simulations. Solid State Ionics 145, 421–427. https://doi.org/10.1016/S0167-2738(01)00939-0

Shimojuku, A., Kubo, T., Kato, T., Yoshino, T., Nishi, M., Nakamura, T., Okazaki, R., Kakazu, Y., 2014. Effects of pressure and temperature on the silicon diffusivity of pyrope-rich garnet. Phys. Earth Planet. Inter. 226, 28–38. https://doi.org/10.1016/j.pepi.2013.11.002

Sholl, D.S.., Steckel, J.A.., 2009. Density functional theory : a practical introduction. Wiley.

Skinner, B., 1956. Physical properties of end-members of the garnet group. Am. Mineral. 41, 428–436.

Spanner, M., Höfling, F., Kapfer, S.C., Mecke, K.R., Schröder-Turk, G.E., Franosch, T., 2016. Splitting of the universality class of anomalous transport in crowded media. Phys. Rev. Lett. 116, 1–6. https://doi.org/10.1103/PhysRevLett.116.060601

Subasinghe, J.L., Ganeshalingam, S., Kuganathan, N., 2022. Computational Study of Crystallography, Defects, Ion Migration and Dopants in Almandine Garnet. Physchem 2, 43–51. https://doi.org/10.3390/physchem2010004

Sun, W., Yoshino, T., Sakamoto, N., Yurimoto, H., 2021. Hydrogen diffusion mechanism in the mantle deduced from H-D interdiffusion in wadsleyite. Earth Planet. Sci. Lett. 561, 116815. https://doi.org/10.1016/j.epsl.2021.116815

Togo, A., 2023. First-principles Phonon Calculations with Phonopy and Phono3py. J. Phys. Soc. Japan 92. https://doi.org/10.7566/JPSJ.92.012001

Unke, O.T., Chmiela, S., Sauceda, H.E., Gastegger, M., Poltavsky, I., Schütt, K.T., Tkatchenko, A., Müller, K.R., 2021. Machine Learning Force Fields. Chem. Rev. 121, 10142–10186. https://doi.org/10.1021/acs.chemrev.0c01111

Verma, A.K., Karki, B.B., 2009. Ab initio investigations of native and protonic point defects in Mg2SiO4 polymorphs under high pressure. Earth Planet. Sci. Lett. 285, 140–149. https://doi.org/10.1016/j.epsl.2009.06.009

Voigtmann, T., Horbach, J., 2009. Double transition scenario for anomalous diffusion in Glass-Forming mixtures. Phys. Rev. Lett. 103, 8–11.

https://doi.org/10.1103/PhysRevLett.103.205901

Wang, H., Zhang, L., Han, J., E, W., 2018. DeePMD-kit: A deep learning package for many-body potential energy representation and molecular dynamics. Comput. Phys. Commun. 228, 178–184. https://doi.org/10.1016/j.cpc.2018.03.016

Wang, L., Zhang, Y., Essene, E.J., 1996. Diffusion of the hydrous component in pyrope. Am. Mineral. 81, 706–718. https://doi.org/10.2138/am-1996-5-618

Xu, L., Mei, S., Dixon, N., Jin, Z., Suzuki, A.M., Kohlstedt, D.L., 2013. Effect of water on rheological properties of garnet at high temperatures and pressures. Earth Planet. Sci. Lett. 379, 158–165. https://doi.org/10.1016/j.epsl.2013.08.002

Xue, X., Kanzaka, M., Turner, D., Loroch, D., 2017. Hydrogen incorporation mechanisms in forsterite: New insights from 1H and 29Si NMR spectroscopy and first-principles calculation. Am. Mineral. 102, 519–536.

Zhang, B., Li, B., Zhao, C., Yang, X., 2019. Large effect of water on Fe–Mg interdiffusion in garnet. Earth Planet. Sci. Lett. 505, 20–29. https://doi.org/10.1016/j.epsl.2018.10.015

Zhang, L., Han, J., Wang, H., Car, R., Weinan, E., 2018a. Deep Potential Molecular Dynamics: A Scalable Model with the Accuracy of Quantum Mechanics. Phys. Rev. Lett. 120, 143001. https://doi.org/10.1103/PhysRevLett.120.143001

Zhang, L., Han, J., Wang, H., Saidi, W.A., Car, R., Weinan, E., 2018b. End-to-end symmetry preserving inter-atomic potential energy model for finite and extended systems. Adv. Neural Inf. Process. Syst. 2018-Decem, 4436–4446.

Zhang, P., Ingrin, J., Depecker, C., Xia, Q., 2015. Kinetics of deuteration in andradite and garnet. Am. Mineral. 100, 1400–1410. https://doi.org/10.2138/am-2015-5149

Zhang, S.B., Northrup, J.E., 1991. Chemical Potential Dependence of Defect Formation Energies in GaAs: Application to Ga self-diffusion. Phys. Rev. Lett. 67, 2339–2342.

Zhang, Y., Liu, X., van Duin, A.C.T., Lu, X., Meijer, E.J., 2024. Development and validation of a general-purpose ReaxFF reactive force field for earth material modeling. J. Chem. Phys. 160. https://doi.org/10.1063/5.0194486

Zhong, X., Wallis, D., Kingsbery, P., John, T., 2024. The effect of aqueous fluid on viscous relaxation of garnet and modification of inclusion pressures after entrapment. Earth Planet. Sci. Lett. 636, 118713. https://doi.org/10.1016/j.epsl.2024.118713

## Figures and tables

Table 1. Physical properties of pyrope garnet at ambient conditions calculated via DFT within the quasi-harmonic approximation (QHA) with the PBE and PBEsol exchange functionals, respectively. To facilitate the comparison, experimental values are quoted from the literature; in each column, the minimum and maximum experimental values are typeset in bold.

| Method | Lattice constant (Å) | Volume (Å$^3$) | Density (kg/m$^3$) | Bulk modulus (GPa) | Isobaric heat capacity $C_p$ (J $K^{-1}$ $mol^{-1}$) | Thermal expansion ($10^{-5}$ $K^{-1}$) | Reference |
|---|---|---|---|---|---|---|---|
| DFT-PBE | 11.621 | 1569.37 | 3412.4 | 169.8 | 341.2 | 2.43 | This study |
| DFT-PBEsol | 11.492 | 1517.84 | 3528.2 | 167.5 | 333.6 | 2.10 | This study |
| Experiments | **11.463** | **1506.15** | **3555.6** | **163.7** | | **3.01** | Milani et al. (2015) |
| | 11.456 | 1503.33 | 3562.3 | | **351.9** | 2.36 | Hartwig and Galkin (2021) |
| | | | | | **334.4** | | Bosenick et al. (1996) |
| | 11.459 | 1504.66 | 3559.2 | **172.6** | | 1.98 | Skinner (1956) |
| | 11.455 | 1503.21 | 3562.6 | | | **1.94** | Bosenick and Geiger (1997) |
| | **11.454** | **1502.8** | **3563.6** | 169.2 | | 2.74 | Du et al. (2015) |

Table 2 Reaction enthalpies of charge-balanced point defects in pyrope garnet calculated via DFT with the PBE and PBEsol exchange functionals; see Section 2.5 for details of the calculation.

| **Defect type** | **Reaction enthalpy (kJ mol$^{-1}$)** | |
|---|---|---|
| **Schottky pair defects (0 K)** | PBE | PBEsol |
| $\mathrm{Mg_{Mg}^{x} + O_{O}^{x} = V_{Mg}'' + V_{O}^{\bullet\bullet} + MgO}$ | 285.5 | 332.8 |
| $\mathrm{Si_{Si}^{x} + 2O_{O}^{x} = V_{Si}'''' + 2V_{O}^{\bullet\bullet} + SiO_2}$ | 741.7 | 863.3 |
| $\mathrm{Al_{Al}^{x} + \frac{3}{2}O_{O}^{x} = V_{Al}''' + \frac{3}{2}V_{O}^{\bullet\bullet} + \frac{1}{2}Al_2O_3}$ | 493.3 | 578.8 |
| **Frenkel pair defects (0 K)** | | |
| $\mathrm{Mg_{Mg}^{x} = V_{Mg}'' + Mg_{I}^{\bullet\bullet}}$ | 565.5 | 597.2 |
| $\mathrm{Si_{Si}^{x} = V_{Si}'''' + Si_{I}^{\bullet\bullet\bullet\bullet}}$ | 882.0 | 949.0 |
| $\mathrm{Al_{Al}^{x} = V_{Al}'' + Al_{I}^{\bullet\bullet}}$ | 720.0 | 781.0 |
| **Protonation defects (300 K)** | | |
| $\mathrm{Mg_{Mg}^{\times} + H_2O = (2H)_{Mg}^{\times} + MgO}$ | 137.6 | 111.4 |
| $\mathrm{Si_{Si}^{\times} + 2H_2O = (4H)_{Si}^{\times} + SiO_2}$ | 65.5 | 14.6 |
| $\mathrm{Al_{Al}^{\times} + \frac{3}{2}H_2O = (3H)_{Al}^{\times} + \frac{1}{2}Al_2O_3}$ | 186.1 | 122.7 |
| **Antisite defects (0 K)** | | |
| $\mathrm{Mg_{Mg}^{\times} + Si_{Si}^{\times} = Si_{Mg}^{\bullet\bullet} + Mg_{Si}''}$ | 530.8 | 566.9 |
| $\mathrm{Mg_{Mg}^{\times} + Al_{Al}^{\times} = Al_{Mg}^{\bullet} + Mg_{Al}'}$ | 194.5 | 210.3 |
| $\mathrm{Si_{Si}^{\times} + Al_{Al}^{\times} = Si_{Al}^{\bullet} + Al_{Si}'}$ | 113.5 | 109.3 |

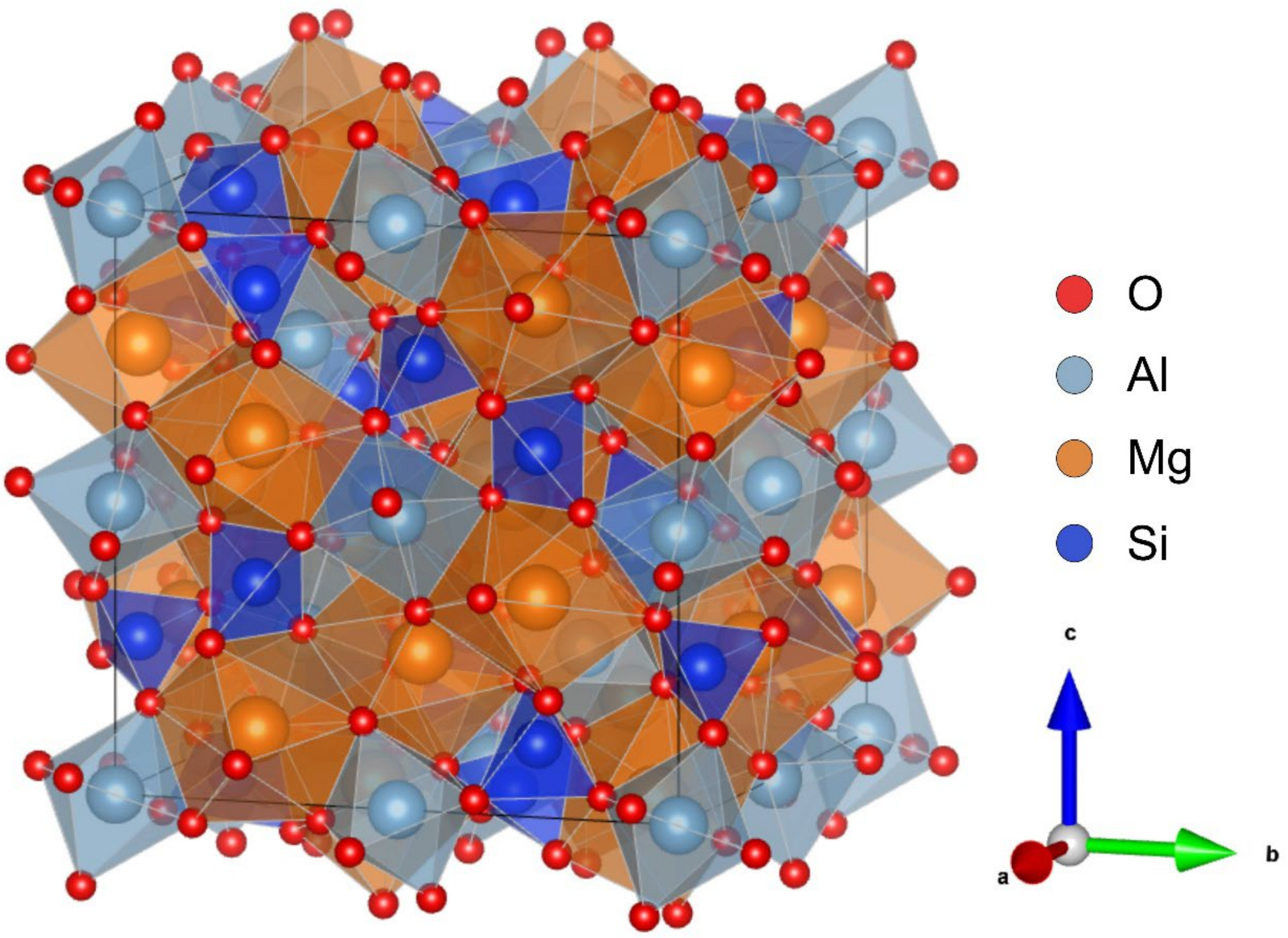

Figure 1. Atomic structure of pyrope garnet. The conventional unit cell shown contains 160 atoms, and the polygons indicate the faces formed by the surrounding O atoms around the Mg, Si or Al atom. The almandine, grossular, and spessartine endmembers correspond to the structures in which Mg atoms are replaced by Fe, Ca, and Mn atoms, respectively.

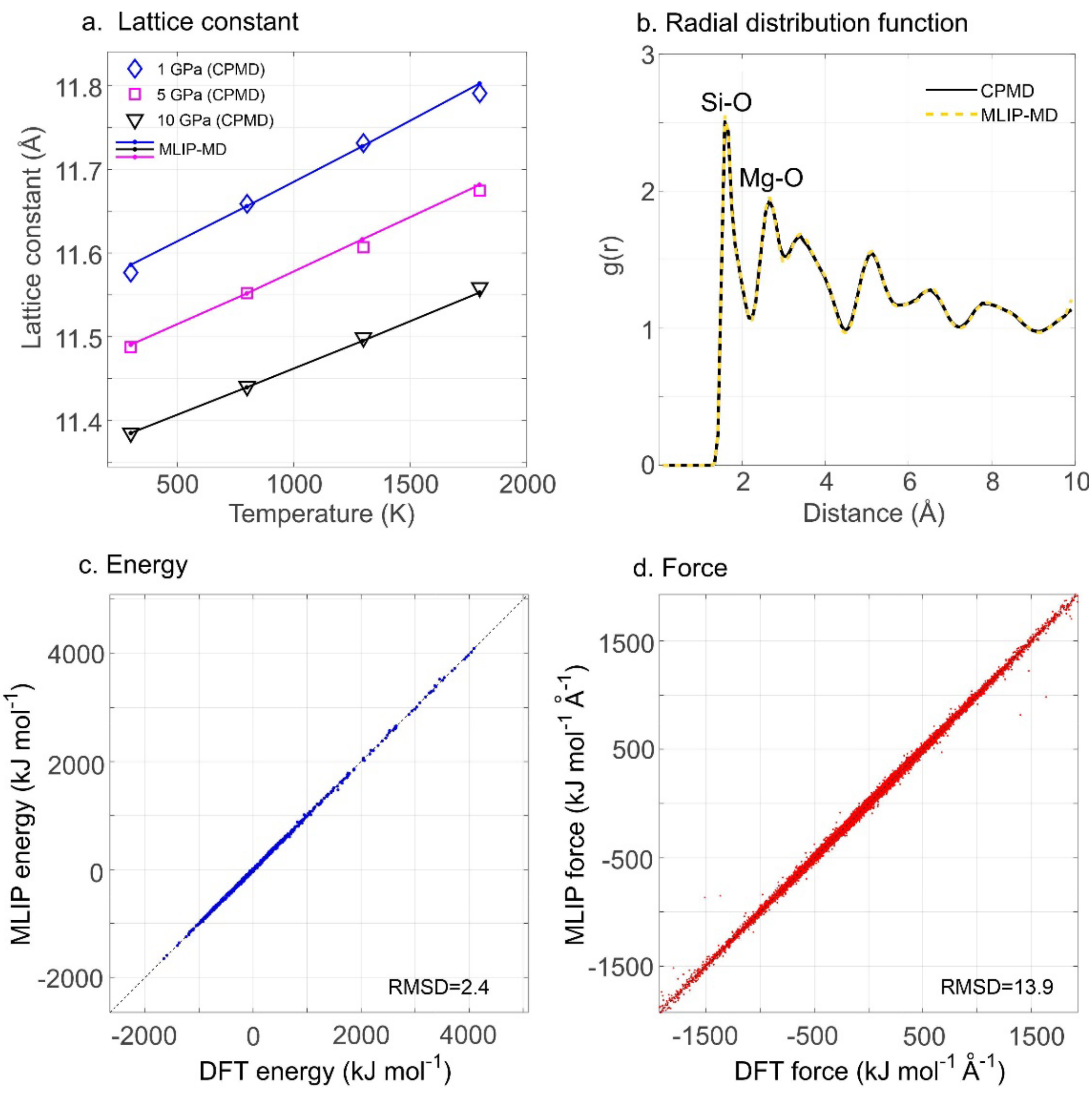

Figure 2. MD simulation results using the machine learned interatomic potential (MLIP). Panel (a) shows the equilibrium lattice constants of a perfect pyrope lattice under high pressure–temperature conditions obtained from CPMD (open symbols) and MLIP-MD (dots, connected by lines) simulations. Panel (b) compares the radial distribution functions obtained from CPMD and MLIP-MD simulations at a temperature of 1800 K and a pressure of 1 GPa. Panel (c) and (d) are parity plots for the energy and atomic force between self-consistent field DFT (x-axis) and MLIP (y-axis) for 5000 random atomic configurations obtained from CPMD simulation. The energy is shifted for illustration.

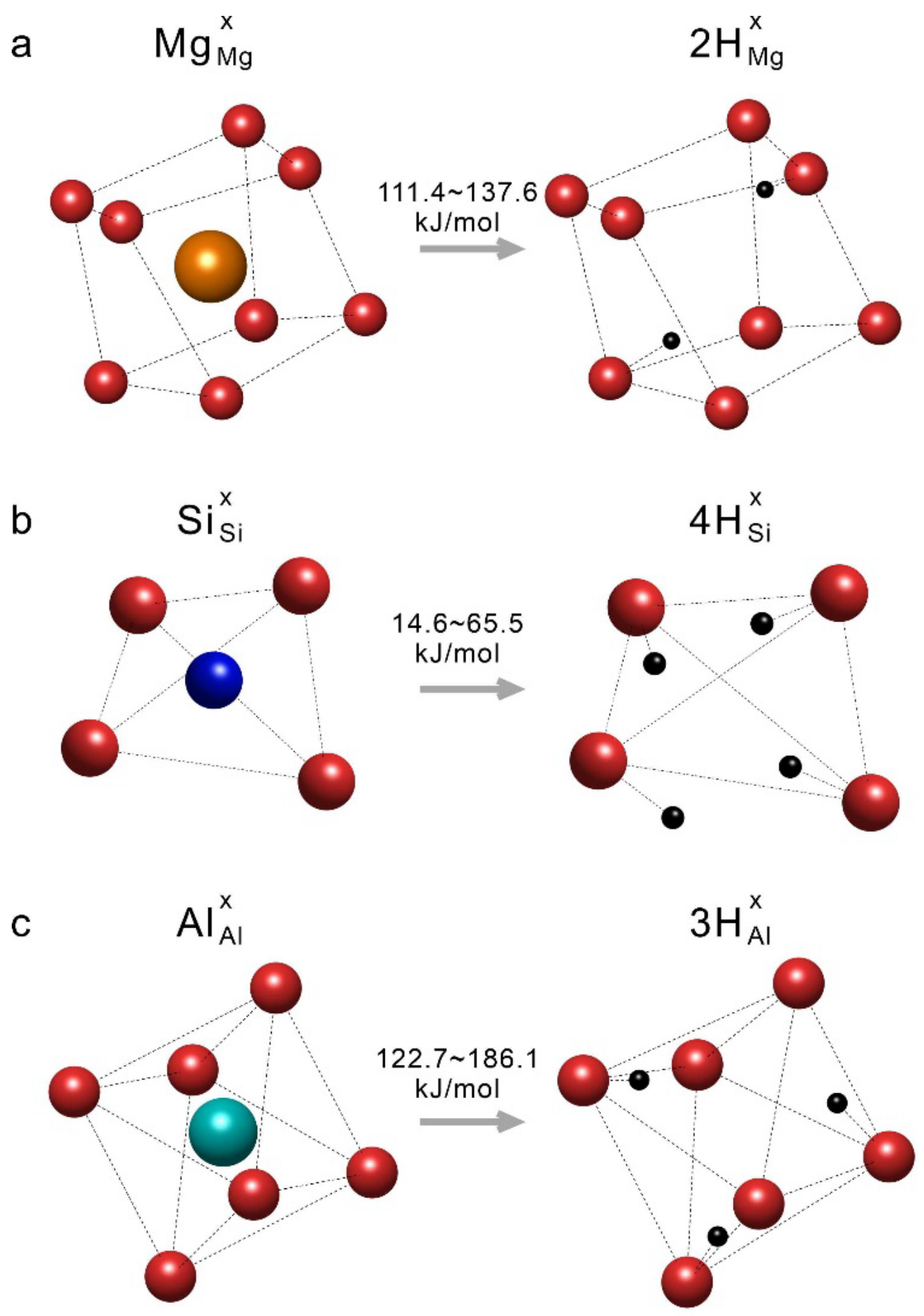

Figure 3. Relaxed hydrogen positions of $(2H)^{\times}_{Mg}$, $(4H)^{\times}_{Si}$, and $(3H)^{\times}_{Al}$ defects in pyrope obtained via DFT calculations. The values above the arrows are estimates of the formation energy of the vacancy upon replacement by hydrogen atoms (black beads); the red beads represent oxygen atoms.

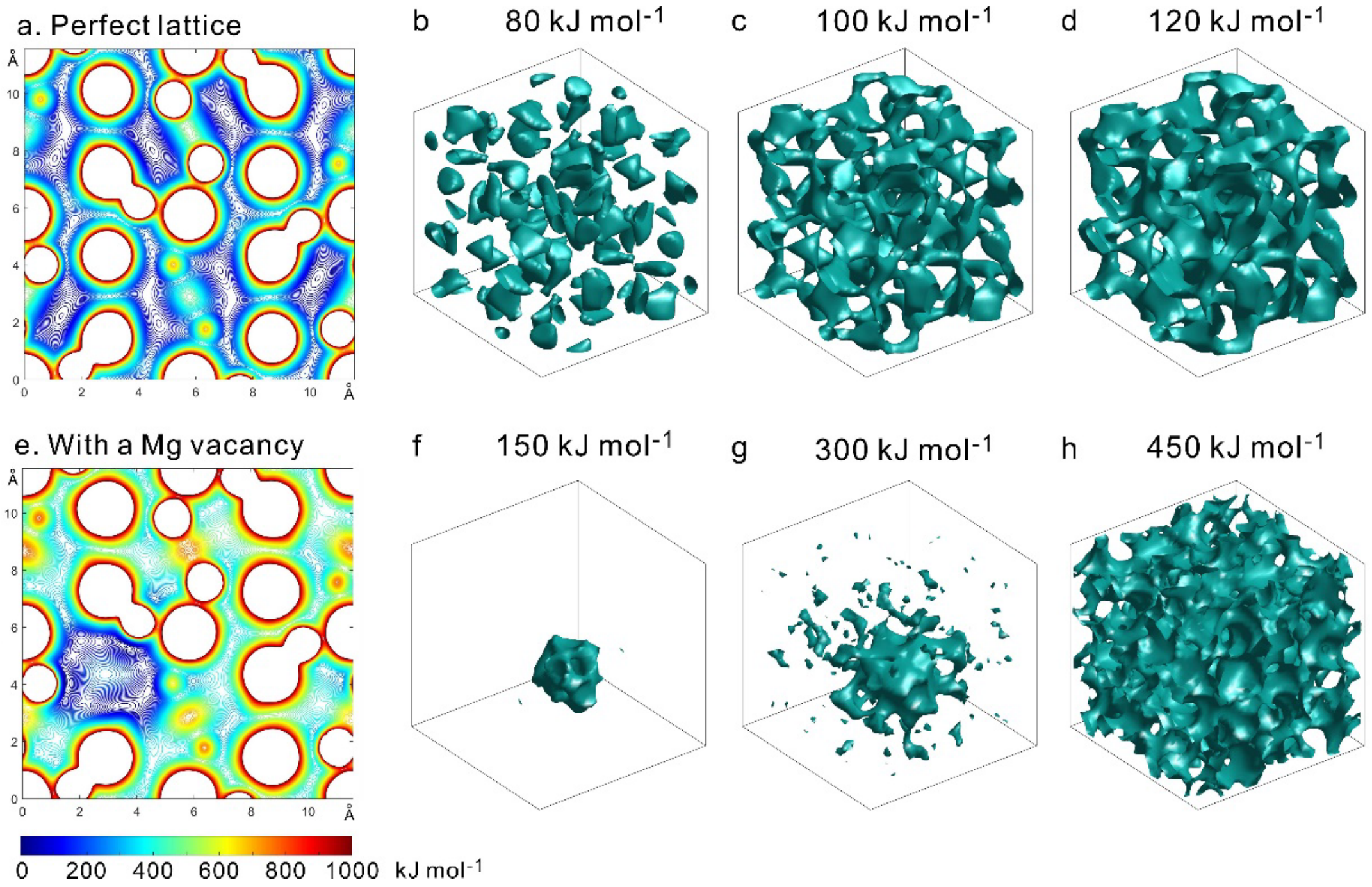

Figure 4. Potential energy surface of a perfect pyrope lattice and a defective lattice with a Mg vacancy. Panel (a) shows a 2D cross-section of the perfect-lattice potential energy surface through the Mg atom; colors encode the energy in kJ $mol^{-1}$. Panels (b) to (d) show 3D isosurfaces at energies of 80, 100, and 120 kJ $mol^{-1}$; the isolated "pockets" at 80 kJ $mol^{-1}$ merge at and above 100 kJ $mol^{-1}$. The lower row of panels, (e) to (h), shows the corresponding potential energy surface in the presence of a Mg vacancy, which appears as a deep potential energy well (blue color) in panel (e).

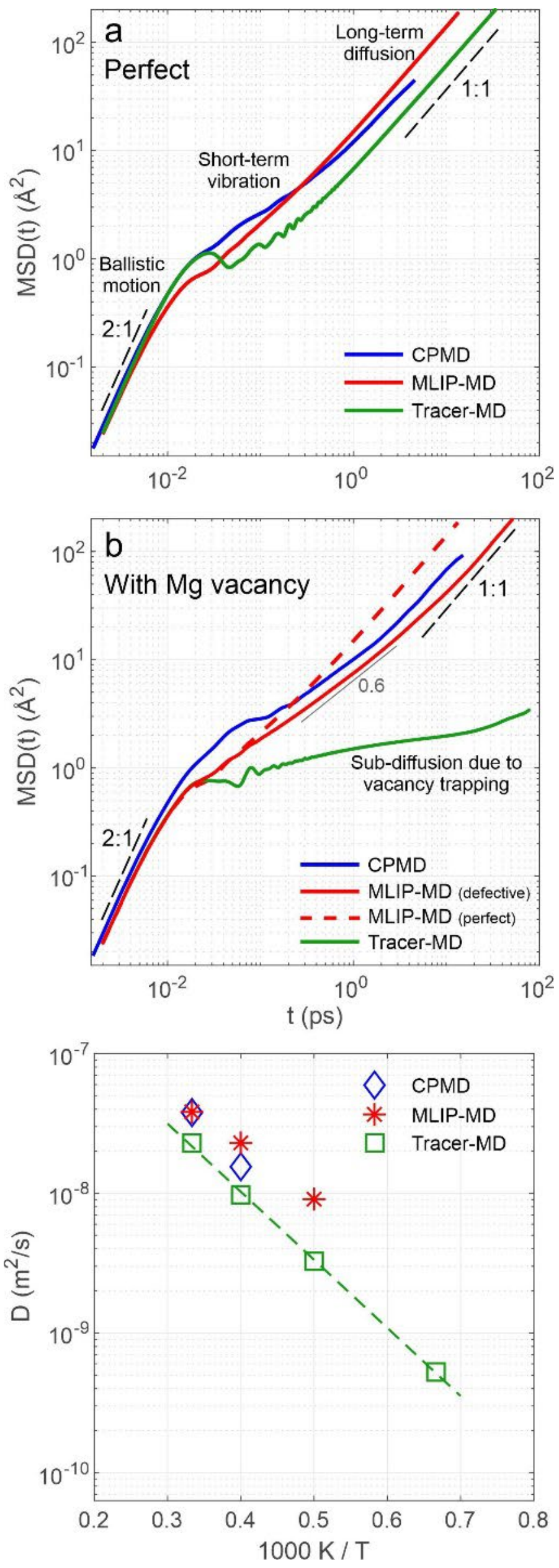

Figure 5. Panels (a) and (b): mean-square displacement (MSD) as a function of the lag time $t$ for the motion of interstitial H atoms in garnet at a temperature of 2500 K and a pressure of 10 GPa for (a) the perfect pyrope lattice and (b) defective garnet with one Mg vacancy, $(2H)_{Mg}^{\times}$. The red dashed line in panel (b) shows the MLIP-MD data from (a) for comparison. Panel (c): Arrhenius plot of the H diffusivity $D$ in the perfect garnet lattice for temperatures between 1500 K and 3000 K. The straight dotted line tests an Arrhenius behavior with an activation energy of $Q = 93$ kJ mol$^{-1}$. In all three panels, the data are from AIMD simulations using either the Car–Parrinello method (CPMD), a machine learned interaction potential (MLIP-MD), or potential energy surface tracers (tracer MD).

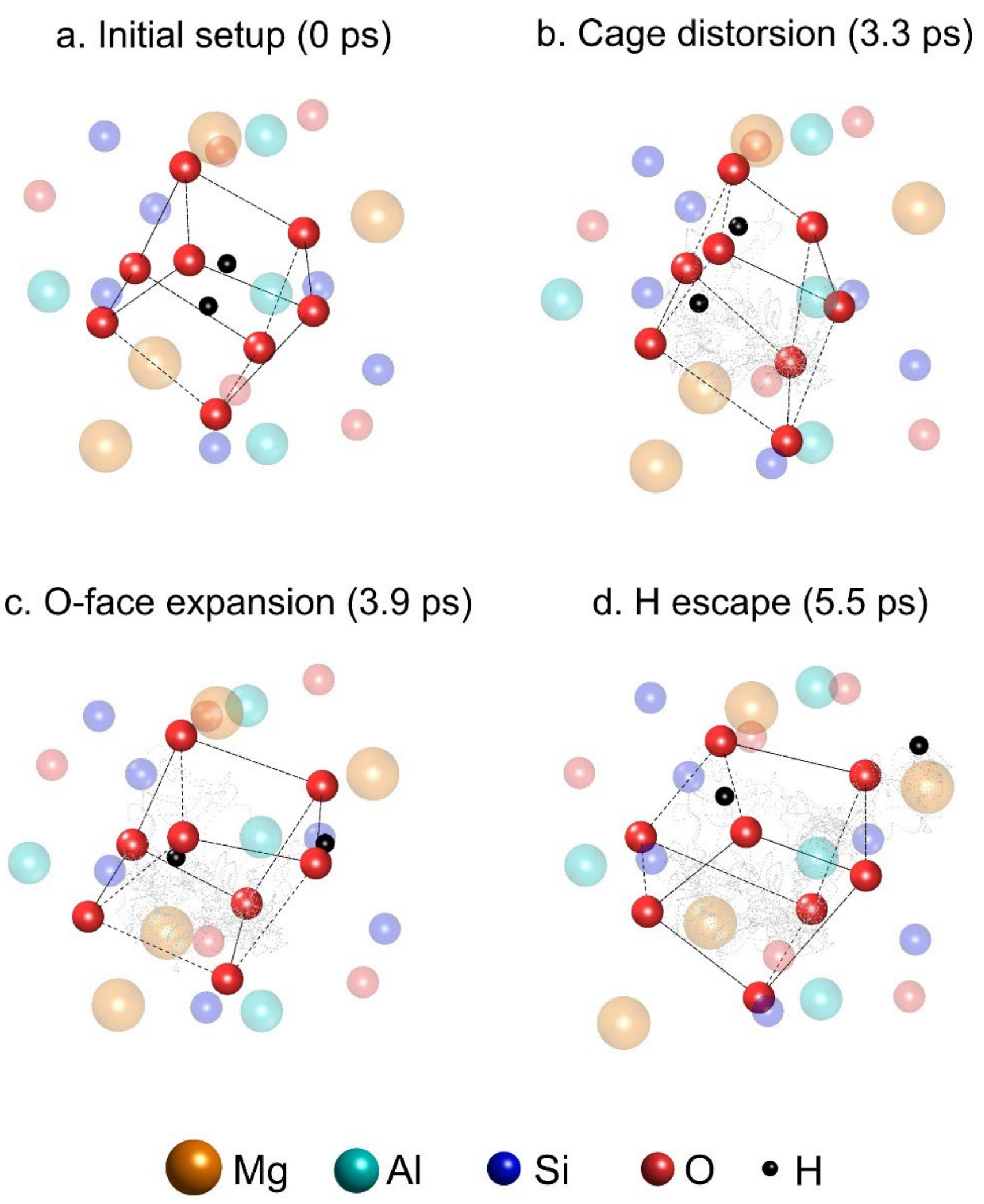

Figure 6. The stages of the H escape dynamics in garnet from a Mg vacancy, depicted by atomic configurations from an MLIP-MD simulation at a temperature of 2500 K. Initially, 2 H atoms are sitting at a vacant Mg site, caged by eight O atoms (panel (a)). (b) Atomic vibrations lead to lattice distortions, but the H atoms are still located within the O cage (panel (b), dashed polyhedron). Eventually, one of the cage faces expands so that the H atom is drawn toward an O atom at the face corner and is about to escape (c). In the end, the H atom moves outside the cage (d). The trajectory of one escaping H atom is shown in light gray, and the atoms outside the O cage are rendered with a transparent color.

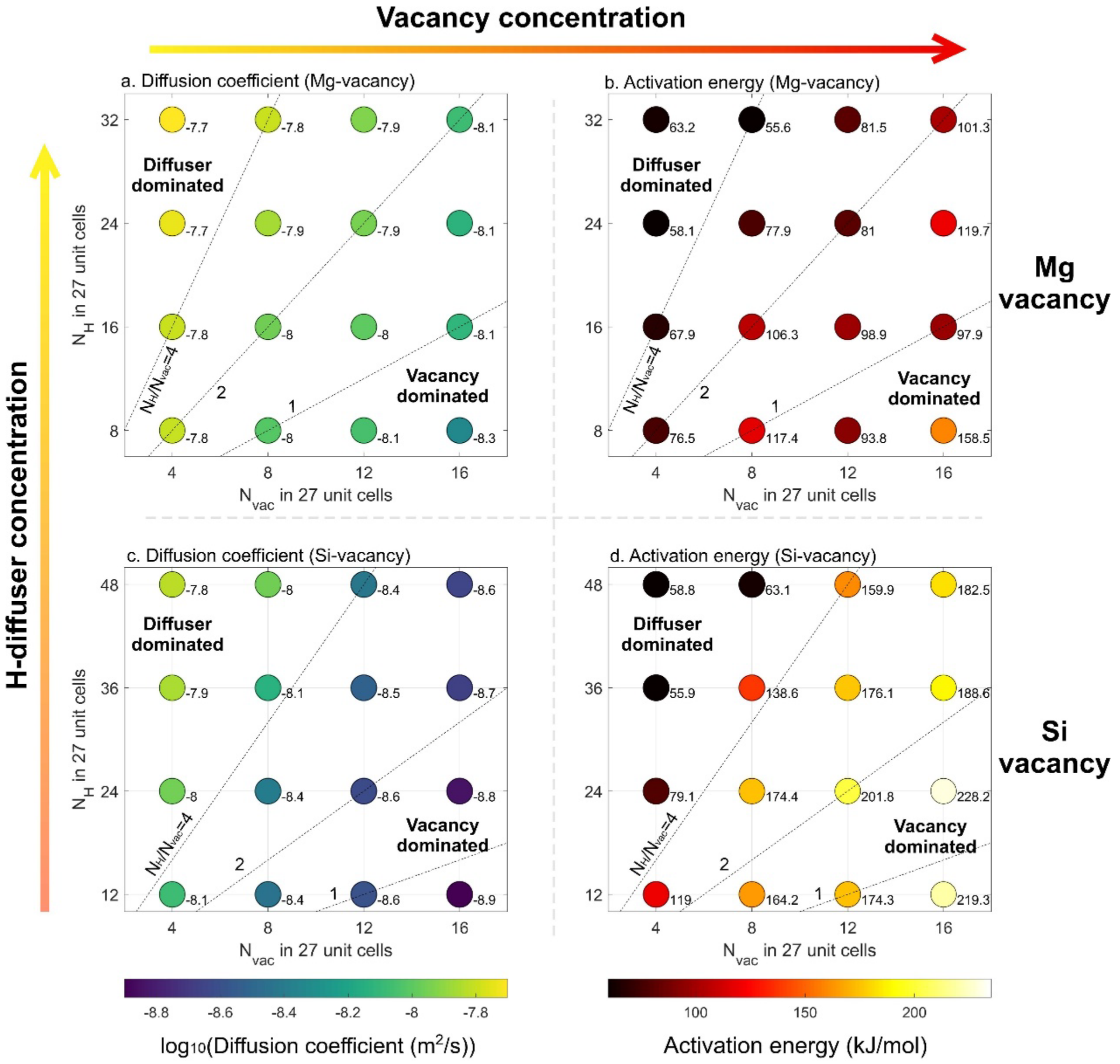

Figure 7. Hydrogen diffusion coefficient (left panels) and corresponding activation energy (right panels) in defect pyrope garnet for different concentrations of Mg vacancies (a, b) and Si vacancies (c, d). Subplots (a) and (c) show the diffusion coefficients calculated at a temperature of 2500 K and a pressure of 10 GPa; subplots (b) and (d) show the activation energy obtained via Arrhenius fits to the diffusion coefficients at 2500 K and 3000 K. Each diffusion coefficient was obtained from a separate, up to 1-ns long MLIP-MD simulation of 3-by-3-by-3 garnet supercells; the latter having a volume of ca. 4.3 $nm^3$ and containing ca. 4300 lattice atoms, $N_H$ interstitial hydrogen atoms, and $N_{\mathrm{vac}}$ vacancies. The ratio $N_H/N_{\mathrm{vac}}$ is plotted as dashed lines.

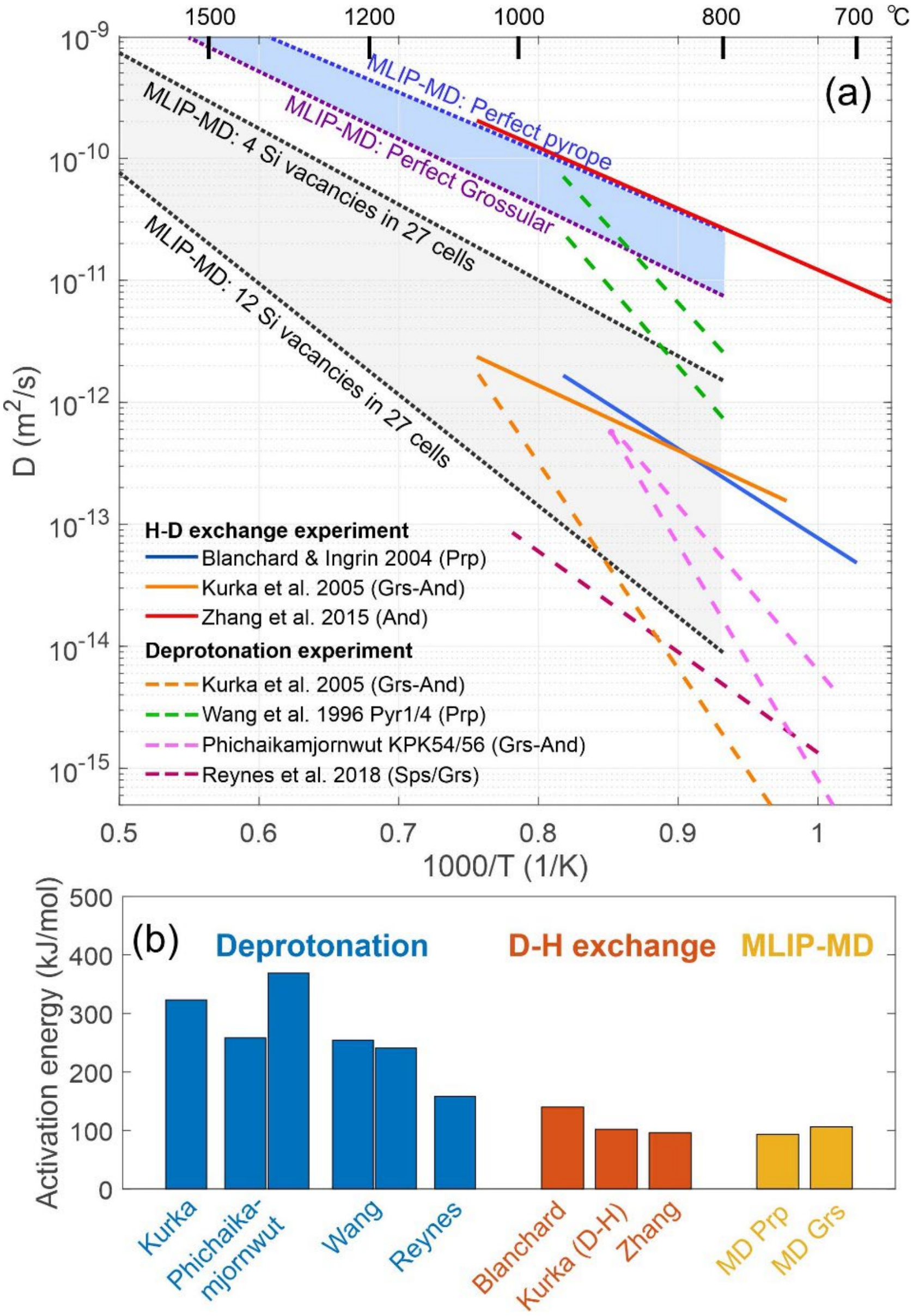

Figure 8. Comparison of (a) the simulated H diffusivity in garnet and (b) the corresponding activation energies with experimental data. In panel (a), the diffusion coefficients from MLIP-MD simulations were extrapolated to 1000~1400 K. The blue shaded area denotes the range for a perfect lattice of pure pyrope (upper blue dashed line) and pure grossular (lower blue dashed line). The gray shaded area denotes the range for the lattice with Si defects from 4 to 12 $V_{Si}^{\times}$ defects in 27 cubic unit cells, which are filled by diffusing H atoms. The experimental data are taken from Blanchard and Ingrin (2004); Kurka et al., (2005); Phichaikamjornwut et al. (2012); Reynes et al., (2018); Zhang et al. (2015) and Wang et al. (1996).